\documentclass[12pt,a4paper]{article}
\usepackage{jheplikemod,ifpdf,tocloft,rotating,amsmath,textcomp,mathrsfs,mathtools,xcolor}
\usepackage{tikz}
\usetikzlibrary{calc,decorations.markings,shapes.geometric}
\usepackage[backgroundcolor=green!5, linecolor=blue!60]{todonotes}

\newcommand{\resetGraphDefaults}{
\definecolor{emphA}{rgb}{0.75,0.0,0.325}
\definecolor{emphB}{rgb}{0.2,0.605,0}
\definecolor{emphC}{rgb}{0,0,0.975}
\definecolor{ndotColor}{rgb}{0.65,0.25,0.25}

\definecolor{optLegColour}{rgb}{0.5,0.5,0.5}
\definecolor{legColour}{rgb}{0.35,0.35,0.35}
\definecolor{legLabelColour}{rgb}{0,0,0}
\definecolor{mhvblue}{rgb}{0.6,0.6,0.7765}
\definecolor{ampgrey}{rgb}{0.9,0.9,0.9}
\def\ampSize{(1*\figScale*7pt)}

\def\figScale{1}
\def\edgeLen{1*\figScale}
\def\fScale{\footnotesize}
\pgfmathsetmacro{\pLen}{\edgeLen/(2*sin(72/2))}
\def\legLen{\edgeLen*0.55}
\def\dotDist{\legLen*0.75}
\def\labelDist{\legLen*1.5}
\def\lineThickness{(1.25pt)}
\def\dotSize{(\figScale*12pt)}

\def\legSpread{4}
\def\dotSize{15*\figScale}
\def\extLegLen{1.12725*0.32*\figScale}
\def\labelDist{0.55*\figScale}
\def\ampSize{(1.1*\figScale*12pt)}

\tikzset{int/.style={black,line width=\lineThickness,line cap=round,rounded corners=0.5pt}}
\tikzset{ddot/.style={fill=black,circle,minimum size=0.35*\dotSize,inner sep=0}}
\tikzset{intH/.style={hred,line width=\lineThickness,line cap=round,rounded corners=0.5pt}}
\tikzset{hdot/.style={fill=hblue,circle,minimum size=0.15*\dotSize,inner sep=0}}
\tikzset{edot/.style={fill=hblue,circle,minimum size=0.15*\dotSize,inner sep=0}}
\tikzset{eddot/.style={fill=black,circle,minimum size=0.15*\dotSize,inner sep=0}}

}
\tikzset{fullamp/.style={coordinate,minimum size=1.5*\ampSize,ball color=black!20,circle,text=white,inner sep=0}}
\tikzset{fullmhv/.style={coordinate,minimum size=0.9*\ampSize,ball color=mhvblue,circle,text=white,inner sep=0}}
\tikzset{fullmhvBar/.style={coordinate,minimum size=0.9*\ampSize,ball color=white,circle,text=white,inner sep=0}}
\tikzset{mhv/.style={fill=mhvblue,circle,draw=black,line width=\lineThickness,minimum size=0.8*\ampSize,text=white,inner sep=0}}
\tikzset{mhvBar/.style={fill=white,circle,draw=black,line width=\lineThickness,minimum size=0.8*\ampSize,text=white,inner sep=0}}
\tikzset{genAmp/.style={fill=ampgrey,circle,draw=black,line width=\lineThickness,minimum size=0.8*\ampSize,text=white,inner sep=0}}

\resetGraphDefaults

\newcommand{\singleLegLabelled}[3]{\node at #1 [ddot]{};\draw[int] #1--($#1+(#2:\extLegLen*1.25)$);
\node at ($#1+(#2:\labelDist*1.35)$)[]{{\footnotesize #3}};}

\newcommand{\leg}[2]{
\fill[legColour] #1--($#1+(#2+\legSpread*3:1.35*\extLegLen)$)--($#1+(#2-\legSpread*3:1.35*\extLegLen)$);
\node at #1 [ddot]{};
}

\newcommand{\dBoxCoords}{\coordinate(v1)at($\figScale*(0,\figScale*0.5)$);\coordinate(v2)at($(v1)+(0:\figScale*1)$);\coordinate(v3)at($(v2)+(-90:\figScale)$);\coordinate(v4)at($(v3)+(180:\figScale)$);\coordinate(v5)at($(v4)+(180:\figScale)$);\coordinate(v6)at($(v5)+(90:\figScale)$);}

\newcommand{\traintrackCoords}{\coordinate(v1)at($\figScale*(0,\figScale*0.5)$);\coordinate(v2)at($(v1)+(0:\figScale*1)$);\coordinate(v3)at($(v2)+(0:\figScale*1)$);\coordinate(v4)at($(v3)+(-90:\figScale)$);\coordinate(v5)at($(v4)+(180:\figScale)$);\coordinate(v6)at($(v5)+(180:\figScale)$);\coordinate(v7)at($(v6)+(180:\figScale)$);\coordinate(v8)at($(v7)+(90:\figScale)$);}

\newcommand{\wheelCoords}{\coordinate(v0)at($\figScale*(0,0)$);\coordinate(v1)at($(v0)+(0:\figScale*1)$);\coordinate(v2)at($(v0)+(60:\figScale*1)$);\coordinate(v3)at($(v0)+(120:\figScale*1)$);\coordinate(v4)at($(v0)+(180:\figScale*1)$);\coordinate(v5)at($(v0)+(240:\figScale*1)$);\coordinate(v6)at($(v0)+(-60:\figScale*1)$);}

\newcommand{\pBoxCoords}{\coordinate(v2)at($\figScale*(1.25,0)$);\coordinate(v1)at($(v2)+(126:\figScale)$);\coordinate(v3)at($(v2)+(-126:\figScale)$);\coordinate(v4)at($(v3)+(-198:\figScale)$);\coordinate(v5)at($(v4)+(180:\figScale)$);\coordinate(v6)at($(v5)+(90:\figScale)$);\coordinate(v7)at($(v6)+(0:\figScale)$);}

\newcommand{\dPentCoords}{\coordinate(v2)at($\figScale*(1.25,0)$);\coordinate(v1)at($(v2)+(126:\figScale)$);\coordinate(v3)at($(v2)+(-126:\figScale)$);\coordinate(v4)at($(v3)+(-198:\figScale)$);\coordinate(v5)at($(v4)+(-162:\figScale)$);\coordinate(v6)at($(v5)+(-234:\figScale)$);\coordinate(v7)at($(v6)+(-306:\figScale)$);\coordinate(v8)at($(v7)+(-378:\figScale)$);}

\newcommand{\pBoxEdges}{\draw[int](v7)--(v1)--(v2)--(v3)--(v4)--(v5)--(v6)--(v7);\draw[int](v7)--(v4);}
\newcommand{\dPentEdges}{\draw[int](v8)--(v1)--(v2)--(v3)--(v4)--(v5)--(v6)--(v7)--(v8);\draw[int](v8)--(v4);}
\newcommand{\dBoxEdges}{\draw[int](v1)--(v2)--(v3)--(v4)--(v5)--(v6)--(v1);\draw[int](v1)--(v4);}

\newcommand{\dBoxAmps}[6]{
\ifthenelse{#1=1}{\node at (v1) [mhvBar] {};}{\ifthenelse{#1=2}{\node at (v1) [mhv] {};}{\node at (v1) [genAmp] {};}};
\ifthenelse{#2=1}{\node at (v2) [mhvBar] {};}{\ifthenelse{#2=2}{\node at (v2) [mhv] {};}{\node at (v2) [genAmp] {};}};
\ifthenelse{#3=1}{\node at (v3) [mhvBar] {};}{\ifthenelse{#3=2}{\node at (v3) [mhv] {};}{\node at (v3) [genAmp] {};}};
\ifthenelse{#4=1}{\node at (v4) [mhvBar] {};}{\ifthenelse{#4=2}{\node at (v4) [mhv] {};}{\node at (v4) [genAmp] {};}};
\ifthenelse{#5=1}{\node at (v5) [mhvBar] {};}{\ifthenelse{#5=2}{\node at (v5) [mhv] {};}{\node at (v5) [genAmp] {};}};
\ifthenelse{#6=1}{\node at (v6) [mhvBar] {};}{\ifthenelse{#6=2}{\node at (v6) [mhv] {};}{\node at (v6) [genAmp] {};}};
}

\newcommand{\pBoxAmps}[7]{
\ifthenelse{#1=1}{\node at (v1) [mhvBar] {};}{\ifthenelse{#1=2}{\node at (v1) [mhv] {};}{\node at (v1) [genAmp] {};}};
\ifthenelse{#2=1}{\node at (v2) [mhvBar] {};}{\ifthenelse{#2=2}{\node at (v2) [mhv] {};}{\node at (v2) [genAmp] {};}};
\ifthenelse{#3=1}{\node at (v3) [mhvBar] {};}{\ifthenelse{#3=2}{\node at (v3) [mhv] {};}{\node at (v3) [genAmp] {};}};
\ifthenelse{#4=1}{\node at (v4) [mhvBar] {};}{\ifthenelse{#4=2}{\node at (v4) [mhv] {};}{\node at (v4) [genAmp] {};}};
\ifthenelse{#5=1}{\node at (v5) [mhvBar] {};}{\ifthenelse{#5=2}{\node at (v5) [mhv] {};}{\node at (v5) [genAmp] {};}};
\ifthenelse{#6=1}{\node at (v6) [mhvBar] {};}{\ifthenelse{#6=2}{\node at (v6) [mhv] {};}{\node at (v6) [genAmp] {};}};
\ifthenelse{#7=1}{\node at (v7) [mhvBar] {};}{\ifthenelse{#7=2}{\node at (v7) [mhv] {};}{\node at (v7) [genAmp] {};}};
}

\newcommand{\dBoxLegs}[6]{
\ifthenelse{#1=1}{\draw[int] (v1)--($(v1)+(90:\extLegLen*1.25)$);}{\ifthenelse{#1=2}{\draw[int] (v1)--($(v1)+(90+20:\extLegLen*1.25)$);\draw[int] (v1)--($(v1)+(90-20:\extLegLen*1.25)$);}{}};
\ifthenelse{#2=1}{\draw[int] (v2)--($(v2)+(45:\extLegLen*1.25)$);}{\ifthenelse{#2=2}{\draw[int] (v2)--($(v2)+(45+45:\extLegLen*1.25)$);\draw[int] (v2)--($(v2)+(45-45:\extLegLen*1.25)$);}{}};
\ifthenelse{#3=1}{\draw[int] (v3)--($(v3)+(-45:\extLegLen*1.25)$);}{\ifthenelse{#3=2}{\draw[int] (v3)--($(v3)+(-45+45:\extLegLen*1.25)$);\draw[int] (v3)--($(v3)+(-45-45:\extLegLen*1.25)$);}{}};
\ifthenelse{#4=1}{\draw[int] (v4)--($(v4)+(-90:\extLegLen*1.25)$);}{\ifthenelse{#4=2}{\draw[int] (v4)--($(v4)+(-90+20:\extLegLen*1.25)$);\draw[int] (v4)--($(v4)+(-90-20:\extLegLen*1.25)$);}{}};
\ifthenelse{#5=1}{\draw[int] (v5)--($(v5)+(-135:\extLegLen*1.25)$);}{\ifthenelse{#5=2}{\draw[int] (v5)--($(v5)+(-135+45:\extLegLen*1.25)$);\draw[int] (v5)--($(v5)+(-135-45:\extLegLen*1.25)$);}{}};
\ifthenelse{#6=1}{\draw[int] (v6)--($(v6)+(135:\extLegLen*1.25)$);}{\ifthenelse{#6=2}{\draw[int] (v6)--($(v6)+(135+45:\extLegLen*1.25)$);\draw[int] (v6)--($(v6)+(135-45:\extLegLen*1.25)$);}{}};
}

\newcommand{\pBoxLegs}[7]{
\ifthenelse{#1=1}{\draw[int] (v1)--($(v1)+(72:\extLegLen*1.25)$);}{\ifthenelse{#1=2}{\draw[int] (v1)--($(v1)+(72+45:\extLegLen*1.25)$);\draw[int] (v1)--($(v1)+(72-20:\extLegLen*1.25)$);}{}};
\ifthenelse{#2=1}{\draw[int] (v2)--($(v2)+(0:\extLegLen*1.25)$);}{\ifthenelse{#2=2}{\draw[int] (v2)--($(v2)+(0+45:\extLegLen*1.25)$);\draw[int] (v2)--($(v2)+(0-45:\extLegLen*1.25)$);}{}};
\ifthenelse{#3=1}{\draw[int] (v3)--($(v3)+(-72:\extLegLen*1.25)$);}{\ifthenelse{#3=2}{\draw[int] (v3)--($(v3)+(-72+45:\extLegLen*1.25)$);\draw[int] (v3)--($(v3)+(-72-45:\extLegLen*1.25)$);}{}};
\ifthenelse{#4=1}{\draw[int] (v4)--($(v4)+(-105:\extLegLen*1.25)$);}{\ifthenelse{#4=2}{\draw[int] (v4)--($(v4)+(-105+20:\extLegLen*1.25)$);\draw[int] (v4)--($(v4)+(-105-20:\extLegLen*1.25)$);}{}};
\ifthenelse{#5=1}{\draw[int] (v5)--($(v5)+(-135:\extLegLen*1.25)$);}{\ifthenelse{#5=2}{\draw[int] (v5)--($(v5)+(-135+45:\extLegLen*1.25)$);\draw[int] (v5)--($(v5)+(-135-45:\extLegLen*1.25)$);}{}};
\ifthenelse{#6=1}{\draw[int] (v6)--($(v6)+(135:\extLegLen*1.25)$);}{\ifthenelse{#6=2}{\draw[int] (v6)--($(v6)+(135+45:\extLegLen*1.25)$);\draw[int] (v6)--($(v6)+(135-45:\extLegLen*1.25)$);}{}};
\ifthenelse{#7=1}{\draw[int] (v7)--($(v7)+(105:\extLegLen*1.25)$);}{\ifthenelse{#7=2}{\draw[int] (v7)--($(v7)+(105+45:\extLegLen*1.25)$);\draw[int] (v7)--($(v7)+(105-45:\extLegLen*1.25)$);}{}}
}

\let\olditemize\itemize\renewcommand{\itemize}{\vspace{-2pt}\olditemize\setlength{\itemsep}{1pt}\setlength{\parskip}{0pt}\setlength{\parsep}{-0pt}}
\let\oldenumerate\enumerate\renewcommand{\enumerate}{\vspace{-4pt}\oldenumerate\setlength{\itemsep}{1pt}\setlength{\parskip}{0pt}\setlength{\parsep}{0pt}}
\makeatletter\renewcommand\section{\addtocontents{toc}{\protect\addvspace{-2.25\p@}}\@startsection {section}{1}{\z@}{-0.0ex \@plus .2ex \@minus 0.2ex}{1ex \@plus.1ex\@minus .5ex}{\normalfont\large\bfseries}}
\renewcommand\subsection{\addtocontents{toc}{\protect\addvspace{-2.5\p@}}\@startsection {subsection}{1}{\z@}{0.5ex \@plus .2ex \@minus 0.2ex}{0.75ex \@plus.1ex\@minus .5ex}{\normalfont\bfseries}}

\DeclareMathOperator*{\mathspan}{\mathrm{span}}
\newcommand{\eq}[1]{\vspace{-0.5pt}\begin{equation}#1\vspace{-0.5pt}\end{equation}}

\newcommand{\fwbox}[2]{\text{\makebox[#1][c]{$\hspace{-150pt}\displaystyle#2\hspace{-150pt}$}}}
\newcommand{\fwboxL}[2]{\text{\makebox[#1][l]{$#2$}}}
\newcommand{\fwboxR}[2]{\text{\makebox[#1][r]{$#2$}}}
\newcommand{\equivR}{\fwbox{14.5pt}{\hspace{-0pt}\fwboxR{0pt}{\raisebox{0.47pt}{\hspace{1.25pt}:\hspace{-4pt}}}=\fwboxL{0pt}{}}}
\newcommand{\equivL}{\fwbox{14.5pt}{\fwboxR{0pt}{}=\fwboxL{0pt}{\raisebox{0.47pt}{\hspace{-4pt}:\hspace{1.25pt}}}}}
\newcommand{\fig}[3]{\raisebox{#1}{\includegraphics[scale=#2]{#3}}}

\newcommand{\bigger}[1]{\raisebox{-0.95pt}{\scalebox{1.25}{$#1$}}}
\newcommand{\mi}{\raisebox{0.75pt}{\scalebox{0.75}{$\hspace{-0.5pt}\,-\,\hspace{-0.5pt}$}}}

\renewcommand{\phi}{\varphi}
\renewcommand{\bar}{\overline}
\renewcommand{\hat}{\widehat}

\newcommand{\ab}[1]{\langle #1\rangle}
\renewcommand{\u}[2]{(\hspace{-0.5pt}#1;\hspace{-1.5pt}#2\hspace{-0.5pt})}
\newcommand{\newcap}{\mathrm{\raisebox{0.75pt}{{$\,\bigcap\,$}}}}

\newcommand{\tncap}{\scalebox{0.8}{$\!\newcap\!$}}
\newcommand{\x}[2]{{\color{black}(}\hspace{-0.85pt}{\color{black}#1}\hspace{-0.25pt}{\color{black}|}\hspace{-0.25pt}{\color{black}#2}\hspace{-0.85pt}{\color{black})}}
\newcommand{\dbar}{\fwboxL{7.2pt}{\raisebox{4.5pt}{\fwboxL{0pt}{\scalebox{1.5}[0.75]{\hspace{1.25pt}\text{-}}}}d}}

\renewcommand{\r}[1]{{\color{hred}#1}}
\renewcommand{\b}[1]{{\color{hblue}#1}}

\definecolor{rindou1}{rgb}{0.4431,0.2862,0.7960}
\definecolor{rindou2}{rgb}{0.0078,0.1215,0.4392}
\definecolor{lapis}{rgb}{0.0.0470,0.2941,0.5568}
\definecolor{emerald}{rgb}{0.31, 0.78, 0.47}
\definecolor{pinegreen}{rgb}{0.0, 0.47, 0.44}
\definecolor{jade}{rgb}{0.0, 0.66, 0.42}
\definecolor{teal}{rgb}{0.0, 0.5, 0.5}
\definecolor{hblue}{rgb}{0,0,0.575}
\definecolor{hred}{rgb}{0.575,0.0,0.225}
\definecolor{hgreen}{rgb}{0.0,0.4,0.2}
\definecolor{hteal}{rgb}{0.0,0.545,0.7451}

\title{\texorpdfstring{{\huge \mbox{Prescriptive Unitarity with}}\\[-6pt]{\huge\mbox{\emph{Elliptic} Leading Singularities}}}{Prescriptive Unitarity with Elliptic Leading Singularities}\\[-0pt]}
\author[a,b]{\vspace{-24pt}Jacob~L.~Bourjaily,}\emailAdd{bourjaily@psu.edu}
\author[a]{Nikhil~Kalyanapuram,}\emailAdd{nkalyanapuram@psu.edu}
\author[a]{Cameron~Langer,}\emailAdd{ckl5552@psu.edu}
\author[a]{Kokkimidis~Patatoukos}\emailAdd{kzp326@psu.edu}
\affiliation[a]{Institute for Gravitation and the Cosmos, Department of Physics,\\Pennsylvania State University, University Park, PA 16802, USA}
\affiliation[b]{Niels Bohr International Academy and Discovery Center, Niels Bohr Institute,\\University of Copenhagen, Blegdamsvej 17, DK-2100, Copenhagen \O, Denmark}
\abstract{
We investigate the consequences of elliptic leading singularities for the unitarity-based representations of two-loop amplitudes in planar, maximally supersymmetric Yang-Mills theory. We show that diagonalizing with respect to these leading singularities ensures that the integrand basis is term-wise pure (suitably generalized, to the elliptic multiple polylogarithms, as necessary). We also investigate an alternative strategy based on diagonalizing a basis of integrands on differential forms; this strategy, while neither term-wise Yangian-invariant nor pure, offers several advantages in terms of complexity. 
}
\preprint{}

\begin{document}
\maketitle
\pagenumbering{roman}

\pagenumbering{roman}

\pagenumbering{arabic}
\vspace{0pt}%
\section{Introduction and Overview}\label{sec:introduction}\vspace{0pt}

Generalized unitarity has proven an extremely powerful framework for the representation of scattering amplitudes at large multiplicity and/or loop order. The basic idea is that any loop \emph{integrand}---a rational differential form on the space of internal loop momenta---can be viewed as an element of a  basis of standardized Feynman loop integrands. Provided the basis of integrands is large enough, it can be used to represent \emph{all} the scattering amplitudes of any theory and spacetime dimension. This idea has a long history (see e.g.\ \cite{Bern:1994zx, Bern:1994cg}); it was formalized and used to famous effect in e.g.\ \cite{Britto:2005fq,Anastasiou:2006jv,Bern:2007ct,Cachazo:2008vp,Berger:2008sj,Abreu:2017ptx,Abreu:2017xsl}, and has been recently refined, generalized, and put to use for many impressive applications (see e.g.\ \cite{Ossola:2007ax,Mastrolia:2010nb,Badger:2013sta,Bern:2015ooa,Bourjaily:2013mma,Bourjaily:2015jna}). 

Among the many advantages of this approach is that the basis of integrands, so long as it is large enough, is sufficient to represent literally all amplitudes (arbitrary multiplicity and states) in a wide class of theories at any loop order. Thus, the integrands in the basis need only be integrated once and for all---reusable for any process of interest. As loop integration has been (and remains) among the hardest problems in perturbative quantum field theory, this is a very important feature. This makes clear the importance of choosing `good' integrands for a basis---the precise measures of which have evolved greatly with time. (Roughly speaking, a good basis would consist of integrands which can be integrated `most easily' or which result in the `simplest' expressions.) 

Another advantage to unitarity is that \emph{coefficients} of particular amplitudes with respect to a basis can be computed in terms of mostly (and often wholly) on-shell data---specifically, on-shell functions \cite{Abreu:2017ptx,Abreu:2017xsl,Bourjaily:2017wjl,Arkani-Hamed:2014bca,Franco:2015rma,Benincasa:2015zna,Benincasa:2016awv,Heslop:2016plj}. When these on-shell functions are leading singularities, they have no internal degrees of freedom. Historically, leading singularities have been defined as maximal co-dimension residues (of polylogarithmic differential forms); more recently, this definition has been modified and generalized to include any full-dimensional compact contour integral of a scattering amplitude integrand \cite{Bourjaily:2020hjv}. Leading singularities have played a key role in the development of our modern understanding of quantum field theory (see e.g.\ \cite{Britto:2004nc,Buchbinder:2005wp,Bern:2007dw,Bern:2007ct,Cachazo:2008dx,Cachazo:2008hp,Spradlin:2008uu,Bourjaily:2011hi,Bourjaily:2015bpz,Bourjaily:2015jna}), and many of the remarkable aspects of scattering amplitudes (their simplicity, and wide range of symmetries) were discovered in this context. For example, the BCFW recursion relations for tree-amplitudes were first discovered in this setting \cite{Britto:2004ap,Britto:2004nj,Britto:2005fq}, as was the infinite-dimensional Yangian symmetry of planar maximally supersymmetric Yang-Mills theory (sYM) \cite{Drummond:2008vq,Alday:2008yw,Drummond:2009fd}, and the correspondence between on-shell functions in sYM and residues of the positroid volume-form in Grassmannian manifolds \cite{Drummond:2010uq,ArkaniHamed:2009dn}.  

When leading singularities are used to determine the coefficients of loop amplitudes with respect to some integrand basis, generalized unitarity becomes a relatively simple problem of linear algebra---matching the `cuts' of field theory against the corresponding cuts of the integrand basis. Until recently, however, it was unclear if \emph{leading} singularities represented \emph{complete} information about perturbative scattering amplitudes even in the simplest theories. The reason for this uncertainty lies in the fact that, for sufficiently large multiplicity and/or loop order, scattering amplitude integrands in most theories are not `$d\!\log$' differential forms \cite{Bloch:2013tra,Bourjaily:2017bsb,Remiddi:2017har,Broedel:2017kkb,Broedel:2017siw,Brown:2010bw,Bourjaily:2018ycu,Bourjaily:2018yfy,Bourjaily:2019hmc} and cannot be characterized by (maximal co-dimension) residues alone. For such cases, the traditional definition of leading singularity becomes incomplete; and the most typical strategy to deal with non-polylogarithmic contributions has been to use the highest co-dimension residues that exist (sub${}^n$-leading singularities), and then use sufficient numbers of off-shell evaluations to match a loop integrand functionally on the remaining degrees of freedom. (Examples of such strategies being used can be found in \cite{Bourjaily:2015jna,Bourjaily:2017wjl}.) The result of this approach, however, has many obvious disadvantages; in particular, it results in representations of amplitudes that (at least term-by-term) involve references to arbitrary choices (for the off-shell evaluation) which can break many of the niceties that scattering amplitudes are known to posses.\\[-5pt]

Before we discuss any concrete examples, it is worth highlighting a conventional difference between this work relative to virtually all existing literature: we have chosen to write all loop integration measures in terms of $\dbar^{4L}\vec{\r{\ell}}$ where $\dbar\equivR d/(2\pi)$ (by analogy to `$\hbar$'). As such, many of our results differ by powers of $(2\pi i)$ relative to those found elsewhere. This choice is motivated by the fact that an integrand normalized to have unit \emph{residues} with respect to the measure $d^{4L}\vec{\r{\ell}}$ will have unit \emph{contour} integrals with respect to $\dbar^{4L}\vec{\r{\ell}}$. As such, most formulae appear identical to other literature; a notable exception, however, is the case of sub-leading singularities, for which our convention requires relative factors of $i$.\\[-5pt]

To illustrate how prescriptive unitarity can work when there are elliptic contributions, consider the elliptic double-box integrand for massless, scalar $\phi^4$-theory in four dimensions: 
\eq{\begin{tikzpicture}[scale=0.8*\figScale,baseline=-3.05,rotate=0]
\dBoxCoords\dBoxEdges
\singleLegLabelled{(v1)}{90}{\b{1}};\singleLegLabelled{(v2)}{90}{\b{2}};\singleLegLabelled{(v2)}{0}{\!\b{3}};\singleLegLabelled{(v3)}{0}{\!\b{4}};\singleLegLabelled{(v3)}{-90}{\b{5}};\singleLegLabelled{(v4)}{-90}{\b{6}};\singleLegLabelled{(v5)}{-90}{\b{7}};\singleLegLabelled{(v5)}{-180}{\b{8}\!};\singleLegLabelled{(v6)}{180}{\b{9}\!};\singleLegLabelled{(v6)}{90}{\b{10}};
\foreach\n in {1,...,6}{\node at (v\n) [ddot] {};};
\end{tikzpicture}\bigger{\;\Leftrightarrow\;}\frac{\dbar^4\r{\ell_1}\,\dbar^4\r{\ell_2}}{\x{\r{\ell_1}}{\b{2}}\x{\r{\ell_1}}{\b{4}}\x{\r{\ell_1}}{\b{6}}\x{\r{\ell_1}}{\r{\ell_2}}\x{\r{\ell_2}}{\b{7}}\x{\r{\ell_2}}{\b{9}}\x{\r{\ell_2}}{\b{1}}}
\equivL\,\mathcal{I}_{\text{db}}^{\phi^4}\,.\label{scalar_double_box_integrand}}
Above, $\x{\r{\ell_i}}{\b{a}}$ represents an ordinary, scalar inverse propagator expressed in dual-momentum coordinates (the details of which we review below) and $\dbar\equivR\frac{d}{2\pi}$. At any rate, (\ref{scalar_double_box_integrand}) is an 8-dimensional (rational) differential form on the space of loop momenta. Taking a contour integral which puts all seven propagators on-shell, however, results in an elliptic differential form on the remaining variable\footnote{There are two solutions to these seven (quadratic) equations; here we write the contour on one of them.}:
\eq{\oint_{\substack{\{|\x{\r{\ell_i}}{\b{a}}|=\epsilon\}\\|\x{\r{\ell_1}}{\r{\ell_2}}|=\epsilon}}\mathcal{I}_{\text{db}}^{\phi^4}=\frac{-i c_y}{\x{\b2}{\b6}\x{\b7}{\b1}\x{\b4}{\b9}}\frac{\dbar\r{\alpha}}{y(\r\alpha)}\label{heptacut_of_the_double_box_integrand}}
where we have used $\r{\alpha}$ to represent the final loop momentum variable, $y^2(\r{\alpha})$ is an irreducible quartic (with coefficients that depend on the momenta of the particles involved), and $c_y$ is a factor introduced to render $y^2(\r\alpha)$ \emph{monic}. The precise details are not important to us now; but it is easy to see that (\ref{heptacut_of_the_double_box_integrand}) represents an elliptic differential form, without any further `residues' on which we may define a traditional leading singularity.

Recently \cite{Bourjaily:2020hjv}, a broader definition of leading singularity has been introduced to include any full-dimensional compact contour-integral of a scattering amplitude integrand. With this new definition, we can in fact define a leading singularity for the double-box integrand by integrating (\ref{heptacut_of_the_double_box_integrand}) over, for example, the $a$-cycle, $\Omega_a$, of the elliptic curve: 
\eq{\oint_{\Omega_a}\oint_{\substack{\{|\x{\r{\ell_i}}{\b{a}}|=\epsilon\}\\|\x{\r{\ell_1}}{\r{\ell_2}}|=\epsilon}}\mathcal{I}_{\text{db}}^{\phi^4}=\frac{2}{\pi}\frac{c_y}{\x{\b2}{\b6}\x{\b7}{\b1}\x{\b4}{\b9}}\frac{1}{\sqrt{(r_3-r_2)(r_4-r_1)}}K[\phi]\equivL\,\mathfrak{e}_a^{\phi^4}\,;\label{phi4_elliptic_leading_singularity}}
here, $\phi$ is a cross-ratio in the roots $r_i$ of the quartic (the details of which are not important to us now). If we wanted to chose a basis of integrands which would be normalized to `match' this leading singularity prescriptively, we would need to normalize the double-box integrand accordingly:
\eq{\fwbox{0pt}{\hspace{-10pt}\mathcal{I}_{\mathrm{db}}^{\phi^4}\mapsto\mathcal{I}_{\mathrm{db}}\equivR\mathcal{I}_{\mathrm{db}}^{\phi^4}/\mathfrak{e}_a^{\phi^4}=
\dbar^4\r{\ell_1}\,\dbar^4\r{\ell_2}\frac{\pi\x{\b2}{\b6}\x{\b7}{\b1}\x{\b4}{\b9}\sqrt{(r_3-r_2)(r_4-r_1)}/K[\phi]}{2c_y\,\x{\r{\ell_1}}{\b{2}}\x{\r{\ell_1}}{\b{4}}\x{\r{\ell_1}}{\b{6}}\x{\r{\ell_1}}{\r{\ell_2}}\x{\r{\ell_2}}{\b{7}}\x{\r{\ell_2}}{\b{9}}\x{\r{\ell_2}}{\b{1}}}.
\label{normalized_basis_integrand_for_scalar_phi4_thy}}}
Thus, a prescriptive representation of the 10-particle scattering amplitude in this theory at two loops would involve a term
\eq{\mathcal{I}_{\text{db}}\times\mathfrak{e}_a^{\phi^4}=\Big(\mathcal{I}_{\mathrm{db}}^{\phi^4}/\mathfrak{e}_a^{\phi^4}\Big)\times\mathfrak{e}_a^{\phi^4}=\mathcal{I}_{\text{db}}^{\phi^4}\,.\label{prescriptive_rep_of_double_box_in_phi4}}

While this example may seem overly trivial (especially considering that the original scalar integrand in (\ref{scalar_double_box_integrand}) is \emph{literally} a term in the Feynman expansion!), the re-writing of it according to prescriptive unitarity according to (\ref{prescriptive_rep_of_double_box_in_phi4}) has a remarkable feature: the now-normalized basis integrand $\mathcal{I}_{\text{db}}$ is in fact dual-conformally invariant and \emph{pure} in the sense defined by the authors of \cite{Broedel:2018qkq}---as such, it is arguably the simplest possible form of the integral (and, presumably, the easiest to integrate). (For a broader discussion of integrand `purity', we suggest the reader consult refs.\ \cite{Broedel:2018qkq,ArkaniHamed:2010gh}; for the present, we merely mention this to emphasize that such integrands, and the differential equations that they satisfy (see e.g.\ \cite{Henn:2013pwa}) have been defined so as to manifest many remarkable properties.)\\

\vspace{0pt}%
\subsection{Organization and Outline}\label{subsec:outline}\vspace{0pt}

In this work, we generalize and expand upon the discussion above to the case of two-loop amplitudes in planar, maximally supersymmetric ($\mathcal{N}\!=\!4$) Yang-Mills theory (sYM). A closed formula for all such amplitude integrands was first derived in \mbox{ref.\ \cite{Bourjaily:2015jna}} (representing an early application of what became known as `prescriptive' unitarity \cite{Bourjaily:2017wjl}), which succeeded despite the presence of elliptic integrals due to carefully-made (but arbitrary) choices for off-shell evaluations in combinations of leading and sub-leading singularities. The resulting representations given in \cite{Bourjaily:2015jna,Bourjaily:2019iqr,Bourjaily:2019gqu} involved terms that were neither Yangian-invariant, nor integrands that were pure. 

In section~\ref{sec:prescriptive_unitarity_review} we review the salient elements of two-loop prescriptive unitarity, as well as the novel generalization of elliptic leading singularities introduced in \cite{Bourjaily:2020hjv}. In section \ref{sec:new_prescriptive_integrands}, the main result of this paper, we revisit the prescriptive unitarity story in the light of our recent work. In particular, we derive two novel representations of amplitudes in planar sYM at two loops, both defined completely prescriptively and unambiguously. The first, described in \mbox{section \ref{subsec:homological_diagonalization}}, involves a prescriptive integrand basis chosen by diagonalization on leading singularities (in the new, broader sense); it is intrinsically homological, and results in a representation of amplitudes that, term-by-term, involves Yangian-invariant coefficients and pure integrals. In \mbox{section \ref{subsec:cohomological_diagonalization}}, we describe an alternative representation based instead on a cohomological diagonalization of the integrand basis. The resulting form is simpler in many ways (especially algebraically), but involves coefficients that are not Yangian-invariant and a basis of loop integrands that are not generally pure.

\newpage\vspace{0pt}%
\section[Review: Prescriptive Integrand Bases for sYM at Two Loops]{Review: Prescriptive Integrand Bases for 2-Loop sYM}\label{sec:prescriptive_unitarity_review}\vspace{0pt}

In this section, we briefly review the ingredients of the representation of two-loop integrands in planar sYM as described in ref.~\cite{Bourjaily:2015jna}. More complete details can be found in \cite{Bourjaily:2015jna,Bourjaily:2017wjl}. For what we need in the following sections, the details of how numerators are chosen for the double-pentagons and pentaboxes will not be critical to us---except for the role played by the double-box integrands as `contact terms' of these basis elements. 

\vspace{0pt}%
\subsection{Bases of Dual-Conformal Integrands: General Structure}\label{subsec:integrand_basis}\vspace{0pt}

A very useful (and arguably accidental) feature of planar integrands at two loops is that a complete and \emph{not over-complete} basis of dual-conformal integrands exists. This is in contrast to one loop or three or more loops, for which dual-conformality apparently requires over-completeness (see e.g.~\cite{Bern:2005iz,Bern:2006ew,ArkaniHamed:2010kv}).

At two loops, a dual-conformal basis can be chosen that consists of three classes of integrands---the double-boxes, pentaboxes, and double-pentagons:
\def\figScale{0.9}
\eq{\left\{\begin{tikzpicture}[scale=1*\figScale,baseline=-3.05,rotate=0]
\dBoxCoords\dBoxEdges
\leg{(v1)}{90};\leg{(v2)}{45};\leg{(v3)}{-45};\leg{(v4)}{-90};\leg{(v5)}{-135};\leg{(v6)}{135};
\end{tikzpicture}
\hspace{5pt},\hspace{5pt}
\begin{tikzpicture}[scale=1*\figScale,baseline=-3.05,rotate=0]
\pBoxCoords\pBoxEdges
\leg{(v1)}{72};\leg{(v2)}{0};\leg{(v3)}{-72};\leg{(v4)}{-105};\leg{(v5)}{-135};\leg{(v6)}{135};\leg{(v7)}{105};
\node at ($(v2)+(180:0.85*\figScale)$) {{\large $ i $}};
\end{tikzpicture}
\hspace{5pt},\hspace{5pt}
\begin{tikzpicture}[scale=1*\figScale,baseline=-3.05,rotate=0]
\dPentCoords
\dPentEdges
\leg{(v1)}{72};\leg{(v2)}{0};\leg{(v3)}{-72};\leg{(v4)}{-90};\leg{(v5)}{-107};\leg{(v6)}{180};\leg{(v7)}{107};\leg{(v8)}{90};
\node at ($(v2)+(180:0.85*\figScale)$) {{\large $ j $}};\node at ($(v6)+(0:0.85*\figScale)$) {{\large $ i $}};
\end{tikzpicture}\right\}\fwboxL{0pt}{\,.}\label{schematic_integrand_basis_topologies}
}
To be clear, these pictures represent the corresponding set of scalar, massless Feynman propagators and the indices $\{i,j\}\!\in\!\{1,2\}$ indicate particular choices of loop-dependent numerators.

All the integrands in our basis can be normalized so-as to be dual-conformal (and when possible, pure). For those integrands {with} \emph{exclusively} residues of maximal co-dimension, they are normalized to have \emph{unit} leading singularities on a choice of such a contour, and made to vanish on all such defining contours for all other integrands in the basis. Most of the integrands, however, have support on double-box sub-topologies which have elliptic structures and therefore cannot be realized as $d\!\log$ differential forms.\\

It is useful to bear in mind that the loop-independent numerators and the overall normalization of the integrands in (\ref{schematic_integrand_basis_topologies}) have considerable flexibility. In particular, even after imposing dual-conformality, the space of possible numerators is relatively large. For the pentabox integrands, dual-conformal, loop independent numerators span a six-dimensional space, which may be decomposed into four contact-term numerators (proportional to one of the four inverse-propagators associated with the edges of the pentagon side of the integrand), and two complementary `top-level' degrees of freedom (schematically indexed by $i\!\in\!\{1,2\}$). 

Graphically, this ambiguity reflects the fact that the double-box topology can be obtained by contracting one of the edges $\{{\color{hblue}a},{\color{hblue}b},{\color{hblue}c},{\color{hblue}d}\}$, 
\eq{\hspace{-4pt}\begin{tikzpicture}[scale=1*\figScale,baseline=-3.05,rotate=0]
\pBoxCoords
\draw[intH](v8)--(v1)--(v2)--(v3)--(v4);\draw[int](v4)--(v5)--(v6)--(v7)--(v8);\draw[int](v8)--(v4);
\leg{(v1)}{72};\leg{(v2)}{0};\leg{(v3)}{-72};\leg{(v4)}{-105};\leg{(v5)}{-135};\leg{(v6)}{135};\leg{(v7)}{105};
\fill[hdot]($($(v7)!.5!(v1)$)+(108:0.2)$) circle (2pt) node[above=1pt] {\normalsize $\b{a} $ };
\fill[hdot]($($(v1)!.5!(v2)$)+(36:0.2)$) circle (2pt) node[above right=1pt] {\normalsize $\b{b}$};
\fill[hdot]($($(v2)!.5!(v3)$)+(-36:0.2)$) circle (2pt) node[below right=1pt] {\normalsize $\b{c}$};
\fill[hdot]($($(v3)!.5!(v4)$)+(-108:0.2)$) circle (2pt) node[below=1pt] {\normalsize $\b{d}$};
\fill[eddot]($($(v4)!.5!(v5)$)+(-90:0.2)$) circle (2pt) node[below=1pt] {\normalsize $\!{e}\,$};
\fill[eddot]($($(v5)!.5!(v6)$)+(180:0.2)$) circle (2pt) node[left=1pt] {\normalsize ${f}$};
\fill[eddot]($($(v6)!.5!(v7)$)+(90:0.2)$) circle (2pt) node[above=1pt] {\normalsize $\!{g}\,$};
\node at ($(v2)+(180:0.85*\figScale)$) {{\large $\b{i}$}};
\end{tikzpicture}\!\bigger{\;\supset}\!\!\left\{\!
\!\!\begin{tikzpicture}[scale=1*\figScale,baseline=-3.05,rotate=0]
\dBoxCoords\draw[intH](v1)--(v2)--(v3)--(v4);\draw[int](v4)--(v5)--(v6)--(v1);\draw[int](v1)--(v4);
\fill[hdot]($($(v1)!.5!(v2)$)+(90:0.2)$) circle (2pt) node[above=2pt] {\normalsize $\b{b}$};
\fill[hdot]($($(v2)!.5!(v3)$)+(0:0.2)$) circle (2pt) node[right=1pt] {\normalsize $\b{c}$};
\fill[hdot]($($(v3)!.5!(v4)$)+(-90:0.2)$) circle (2pt) node[below=1pt] {\normalsize $\b{d}$};
\fill[eddot]($($(v4)!.5!(v5)$)+(-90:0.2)$) circle (2pt) node[below=1pt] {\normalsize $\!{e}\,$};
\fill[eddot]($($(v5)!.5!(v6)$)+(180:0.2)$) circle (2pt) node[left=1pt] {\normalsize ${f}$};
\fill[eddot]($($(v6)!.5!(v1)$)+(90:0.2)$) circle (2pt) node[above=1pt] {\normalsize $\!{g}\,$};
\leg{(v1)}{90};\leg{(v2)}{45};\leg{(v3)}{-45};\leg{(v4)}{-90};\leg{(v5)}{-135};\leg{(v6)}{135};
\end{tikzpicture},\!\!
\begin{tikzpicture}[scale=1*\figScale,baseline=-3.05,rotate=0]
\dBoxCoords\draw[intH](v1)--(v2)--(v3)--(v4);\draw[int](v4)--(v5)--(v6)--(v1);\draw[int](v1)--(v4);
\fill[hdot]($($(v1)!.5!(v2)$)+(90:0.2)$) circle (2pt) node[above=2pt] {\normalsize $\b{a}$};
\fill[hdot]($($(v2)!.5!(v3)$)+(0:0.2)$) circle (2pt) node[right=1pt] {\normalsize $\b{c}$};
\fill[hdot]($($(v3)!.5!(v4)$)+(-90:0.2)$) circle (2pt) node[below=1pt] {\normalsize $\b{d}$};
\fill[eddot]($($(v4)!.5!(v5)$)+(-90:0.2)$) circle (2pt) node[below=1pt] {\normalsize $\!{e}\,$};
\fill[eddot]($($(v5)!.5!(v6)$)+(180:0.2)$) circle (2pt) node[left=1pt] {\normalsize ${f}$};
\fill[eddot]($($(v6)!.5!(v1)$)+(90:0.2)$) circle (2pt) node[above=1pt] {\normalsize $\!{g}\,$};
\leg{(v1)}{90};\leg{(v2)}{45};\leg{(v3)}{-45};\leg{(v4)}{-90};\leg{(v5)}{-135};\leg{(v6)}{135};
\end{tikzpicture},\!\!
\begin{tikzpicture}[scale=1*\figScale,baseline=-3.05,rotate=0]
\dBoxCoords\draw[intH](v1)--(v2)--(v3)--(v4);\draw[int](v4)--(v5)--(v6)--(v1);\draw[int](v1)--(v4);
\fill[hdot]($($(v1)!.5!(v2)$)+(90:0.2)$) circle (2pt) node[above=2pt] {\normalsize $\b{a}$};
\fill[hdot]($($(v2)!.5!(v3)$)+(0:0.2)$) circle (2pt) node[right=1pt] {\normalsize $\b{b}$};
\fill[hdot]($($(v3)!.5!(v4)$)+(-90:0.2)$) circle (2pt) node[below=1pt] {\normalsize $\b{d}$};
\fill[eddot]($($(v4)!.5!(v5)$)+(-90:0.2)$) circle (2pt) node[below=1pt] {\normalsize $\!{e}\,$};
\fill[eddot]($($(v5)!.5!(v6)$)+(180:0.2)$) circle (2pt) node[left=1pt] {\normalsize ${f}$};
\fill[eddot]($($(v6)!.5!(v1)$)+(90:0.2)$) circle (2pt) node[above=1pt] {\normalsize $\!{g}\,$};
\leg{(v1)}{90};\leg{(v2)}{45};\leg{(v3)}{-45};\leg{(v4)}{-90};\leg{(v5)}{-135};\leg{(v6)}{135};
\end{tikzpicture},\!\!
\begin{tikzpicture}[scale=1*\figScale,baseline=-3.05,rotate=0]
\dBoxCoords\draw[intH](v1)--(v2)--(v3)--(v4);\draw[int](v4)--(v5)--(v6)--(v1);\draw[int](v1)--(v4);
\fill[hdot]($($(v1)!.5!(v2)$)+(90:0.2)$) circle (2pt) node[above=2pt] {\normalsize $\b{a}$};
\fill[hdot]($($(v2)!.5!(v3)$)+(0:0.2)$) circle (2pt) node[right=1pt] {\normalsize $\b{b}$};
\fill[hdot]($($(v3)!.5!(v4)$)+(-90:0.2)$) circle (2pt) node[below=2pt] {\normalsize $\b{c}$};
\fill[eddot]($($(v4)!.5!(v5)$)+(-90:0.2)$) circle (2pt) node[below=1pt] {\normalsize $\!{e}\,$};
\fill[eddot]($($(v5)!.5!(v6)$)+(180:0.2)$) circle (2pt) node[left=1pt] {\normalsize ${f}$};
\fill[eddot]($($(v6)!.5!(v1)$)+(90:0.2)$) circle (2pt) node[above=1pt] {\normalsize $\!{g}\,$};
\leg{(v1)}{90};\leg{(v2)}{45};\leg{(v3)}{-45};\leg{(v4)}{-90};\leg{(v5)}{-135};\leg{(v6)}{135};
\end{tikzpicture}\hspace{-3pt}
\right\}\nonumber}
Algebraically, this can be understood by decomposing the vector space of numerators into the following basis, 
\eq{[\r{\ell_1}]\equivR\mathspan\!\left\{\x{\r{\ell_1}}{{\color{hblue}\mathcal{Y}^1}},\x{\r{\ell_1}}{{\color{hblue}\mathcal{Y}^2}},\x{\r{\ell_1}}{\b{a}},\x{\r{\ell_1}}{\b{b}},\x{\r{\ell_1}}{\b{c}},\x{\r{\ell_1}}{\b{d}}\right\},}
where ${\color{hblue}\mathcal{Y}^{1,2}}$ are the solutions to the quadruple cut $\x{\r{\ell_1}}{\b{a}}=\x{\r{\ell_1}}{\b{b}}=\x{\r{\ell_1}}{\b{c}}=\x{\r{\ell_1}}{\b{d}}=0$, whose precise form does not concern us here. Thus, each pentabox integrand,
\eq{\begin{tikzpicture}[scale=1*\figScale,baseline=-3.05,rotate=0]
\pBoxCoords
\draw[intH](v8)--(v1)--(v2)--(v3)--(v4);\draw[int](v4)--(v5)--(v6)--(v7)--(v8);\draw[int](v8)--(v4);
\leg{(v1)}{72};\leg{(v2)}{0};\leg{(v3)}{-72};\leg{(v4)}{-105};\leg{(v5)}{-135};\leg{(v6)}{135};\leg{(v7)}{105};
\fill[hdot]($($(v7)!.5!(v1)$)+(108:0.2)$) circle (2pt) node[above=1pt] {\normalsize $\b{a}$};
\fill[hdot]($($(v1)!.5!(v2)$)+(36:0.2)$) circle (2pt) node[above right=1pt] {\normalsize $\b{b}$};
\fill[hdot]($($(v2)!.5!(v3)$)+(-36:0.2)$) circle (2pt) node[below right=1pt] {\normalsize $\b{c}$};
\fill[hdot]($($(v3)!.5!(v4)$)+(-108:0.2)$) circle (2pt) node[below=1pt] {\normalsize $\b{d}$};
\fill[eddot]($($(v4)!.5!(v5)$)+(-90:0.2)$) circle (2pt) node[below=1pt] {\normalsize $\!{e}\,$};
\fill[eddot]($($(v5)!.5!(v6)$)+(180:0.2)$) circle (2pt) node[left=1pt] {\normalsize ${f}$};
\fill[eddot]($($(v6)!.5!(v7)$)+(90:0.2)$) circle (2pt) node[above=1pt] {\normalsize $\!{g}\,$};
\node at ($(v2)+(180:0.85*\figScale)$) {{\large $\b{i}$}};
\end{tikzpicture}\bigger{\;\Leftrightarrow\;}\dbar^4\r{\ell_1}\,\dbar^4\r{\ell_2}\frac{\mathbf{n}_0^{i}\x{\r{\ell_1}}{\b{\mathcal{Y}^i}}{+}\mathbf{n}^i_{\b{a}}\x{\r{\ell_1}}{\b{a}}{+}\mathbf{n}^i_{\b{b}}\x{\r{\ell_1}}{\b{b}}{+}\mathbf{n}^i_{\b{c}}\x{\r{\ell_1}}{\b{c}}{+}\mathbf{n}^i_{\b{d}}\x{\r{\ell_1}}{\b{d}}}{\x{\r{\ell_1}}{\b{a}}\x{\r{\ell_1}}{\b{b}}\x{\r{\ell_1}}{\b{c}}\x{\r{\ell_1}}{\b{d}}\x{\r{\ell_1}}{{\ell_2}}\x{{\ell_2}}{{e}}\x{{\ell_2}}{{f}}\x{{\ell_2}}{{g}}}}
is not fully-specified until the four (loop-independent, but kinematic-dependent) `constants' $\mathbf{n}^i$ have been specified. The top-level normalization, $\mathbf{n}_0^{i}$ is determined by requiring that (some combination of) the \emph{polylogarithmic} leading-singularity which encircles the eight Feynman propagators of the pentabox is unity.

\vspace{0pt}%
\subsection{Triangular Structure of Basis Integrands' Contact-Terms}\label{subsec:triangular_structure}\vspace{0pt}

In prescriptive unitarity, the basis of integrands is chosen to be diagonal with respect to some choice of leading singularities or `cuts' (or, as we will explore later, with respect to forms). To see how this works, consider the double-pentagon integrands. Among all the integrands in the basis (\ref{schematic_integrand_basis_topologies}), only the double-pentagons have support on the so-called `kissing-box' leading singularities of field theory:
\def\extLegLen{0.37*\figScale}
\eq{\begin{tikzpicture}[scale=1*\figScale,baseline=-3.05,rotate=0]
\coordinate(v4)at($\figScale*(0,0)$);\coordinate(v1)at($(v4)+(45:\figScale)$);\coordinate(v2)at($(v1)+(-45:\figScale)$);\coordinate(v3)at($(v4)+(-45:\figScale)$);\coordinate(v5)at($(v4)+(-135:\figScale)$);\coordinate(v6)at($(v5)+(-225:\figScale)$);\coordinate(v7)at($(v4)+(135:\figScale)$);\coordinate(v8)at($(v7)+(-378:\figScale)$);
\draw[int](v4)--(v1)--(v2)--(v3)--(v4)--(v5)--(v6)--(v7)--(v4);
\leg{(v1)}{90};\leg{(v2)}{0};\leg{(v3)}{-90};\leg{(v4)}{-90};\leg{(v4)}{90};\leg{(v5)}{-90};\leg{(v6)}{180};\leg{(v7)}{90};
\node at ($(v2)+(180:0.71*\figScale)$) {{\large $j$}};\node at ($(v6)+(0:0.71*\figScale)$) {{\large $i$}};
\foreach\n in {1,...,7}{\node at (v\n) [genAmp] {};};
\end{tikzpicture}\fwboxL{0pt}{\,.}\label{kissing_box_ls}}
Here, the indices $\{i,j\}\!\in\!\{1,2\}$ label the two solutions (for each loop separately) to the cut-equations which put these eight propagators on-shell. Thus, we choose the $2\!\times\!2$ non-contact-term degrees of freedom of the double-pentagon integrands to be unit on the corresponding contour integrals (and to vanish on the other solutions). As no other integrands in the basis have support on these cuts---there being no other integrands with a corresponding subset of propagators---we are ensured to match field theory on all these contour integrals manifestly, regardless of how the rest of the integrand basis is chosen. 

However, notice that matching the kissing-box contours (\ref{kissing_box_ls}) in field theory using double-pentagon integrands would be safe \emph{regardless} of the particular choices for the contact terms of the double pentagon. Thus, there remains a $(6^2\mi4\!=)32$-dimensional space of possible double-pentagon integrands for our basis even after the top-level degrees of freedom have been chosen, reflecting the 4 contact term ambiguities for each loop's numerator. Of these, 16 contact-term degrees of freedom will be fixed by the requirement that the double-pentagon integrands vanish on the leading singularities upon which the pentaboxes get normalized; the remaining 16 correspond to (potentially elliptic) double-boxes, about which we will have much more to say.\\

The pentabox integrands, in turn, have a bit more subtlety in their definition. For one thing, because the pentabox integrands have only \emph{two} non-contact-term degrees of freedom in their numerators, they cannot be used to match all \emph{four} pentabox leading-singularities:
\eq{\begin{tikzpicture}[scale=1*\figScale,baseline=-3.05,rotate=0]
\pBoxCoords\pBoxEdges
\leg{(v1)}{72};\leg{(v2)}{0};\leg{(v3)}{-72};\leg{(v4)}{-105};\leg{(v5)}{-135};\leg{(v6)}{135};\leg{(v7)}{105};
\node at ($(v2)+(180:0.85*\figScale)$) {{\large $j$}};
\node at ($(v2)+(180:2.05*\figScale)$) {{\large $i$}};
\foreach\n in {1,...,7}{\node at (v\n) [genAmp] {};};
\end{tikzpicture}\label{pentabox_ls}}
where, as before $\{i,j\}\!\in\!\{1,2\}$ label the particular solutions to the cut-equations which put the eight propagators on-shell. Relatedly, it is not possible to make the double-pentagon integrands vanish on all pentabox contours. This apparent tension is in fact fairly trivial and easy to resolve: we merely need to make \emph{some} choice of two of the pentabox leading singularities (or two independent combinations thereof), on which to normalize the pentabox integrands, and then require that the double-pentagons vanish on each of these contours. Once this is done, all four of the pentabox leading singularities will be matched in the basis: two, manifestly by the pentaboxes, and the complementary pair matched indirectly by residue theorems involving the double-pentagon integrands. This may seem magical, but follows necessarily from the completeness of the integrand basis.\\

What remains open, however, are the questions of the double-boxes: how should they be normalized, and how should the double-box contact terms of the pentaboxes and double-pentagons be defined? For all the double-box integrands which \emph{do} support residues of maximal co-dimension (that is, any which involves at least one three-particle vertex), the strategy above may be iterated once more without further subtlety, using a `composite' leading singularity. (For amplitudes with fewer than ten particles, \emph{all} double-boxes have support on such additional `cuts'; and a complete basis can be made polylogarithmic and `pure' in this way.) 

For double-box integrands \emph{without} (traditional) leading singularities, however, the description above falls short (or at least, is incomplete). We will not review how this question was resolved by the authors of ref.\ \cite{Bourjaily:2015jna,Bourjaily:2017wjl,Bourjaily:2020qca}, in part because we will find more elegant solutions here. For the sake of our current investigation, however, let us assume that a spanning set of maximal co-dimension, polylogarithmic contours have been chosen to define the top-level degrees of freedom of the pentabox and double-pentagon integrands; as such, we may take for granted that all the pentabox and kissing-box leading singularities of field theory are matched by these integrands in the basis---leaving only the question of matching the sub-leading singularities associated with non-polylogarithmic double-boxes. That is, we need only to address the double-box integrands---to determine their normalizations, their coefficients in field theory, and how to `diagonalize' the pentaboxes and double-pentagons with respect to these choices. Let us therefore consider these integrands in some detail.

\vspace{0pt}%
\subsection{Elliptic Double-Box Integrands}\label{subsec:double_box_integrands}\vspace{0pt}

Up to a normalizing factor denoted $\mathbf{n}_{\text{db}}$, the double-box integrands in the DCI-power-counting basis may be defined to be
\eq{\mathcal{I}_{\text{db}}\bigger{\;\Leftrightarrow\;}\begin{tikzpicture}[scale=1*\figScale,baseline=-3.05,rotate=0]
\dBoxCoords\dBoxEdges
\leg{(v1)}{90};\leg{(v2)}{45};\leg{(v3)}{-45};\leg{(v4)}{-90};\leg{(v5)}{-135};\leg{(v6)}{135};
\foreach\n in {1,...,6}{\node at (v\n) [ddot] {};};
\fill[hdot]($($(v1)!.5!(v2)$)+(90:0.2)$) circle (2pt) node[above=2pt] {\normalsize $\b{a}$};
\fill[hdot]($($(v2)!.5!(v3)$)+(0:0.2)$) circle (2pt) node[right=1pt] {\normalsize $\b{b}$};
\fill[hdot]($($(v3)!.5!(v4)$)+(-90:0.2)$) circle (2pt) node[below=2pt] {\normalsize $\b{c}$};
\fill[hdot]($($(v4)!.5!(v5)$)+(-90:0.2)$) circle (2pt) node[below=1pt] {\normalsize $\!\b{e}\,$};
\fill[hdot]($($(v5)!.5!(v6)$)+(180:0.2)$) circle (2pt) node[left=1pt] {\normalsize $\b{f}$};
\fill[hdot]($($(v6)!.5!(v1)$)+(90:0.2)$) circle (2pt) node[above=1pt] {\normalsize $\!\b{g}\,$};
\node at ($($(v2)!.5!(v3)$)+(180:0.4)$)  {{\normalsize $\r{\ell_1}$}};
\node at ($($(v1)!.5!(v4)$)+(180:0.4)$)  {{\normalsize $\r{\ell_2}$}};
\end{tikzpicture}\bigger{\;\Leftrightarrow\;}\dbar^4\r{\ell_1}\,\dbar^4\r{\ell_2}\frac{\mathbf{n}_{\text{db}}\,\x{\b{a}}{\b{c}}\x{\b{d}}{\b{f}}\x{\b{b}}{\b{e}}}{\x{\r{\ell_1}}{\b{a}}\x{\r{\ell_1}}{\b{b}}\x{\r{\ell_1}}{\b{c}}\x{\r{\ell_1}}{\r{\ell_2}}\x{\r{\ell_2}}{\b{d}}\x{\r{\ell_2}}{\b{e}}\x{\r{\ell_2}}{\b{f}}}
\,.\label{scalar_double_box_integrand_general}}
Here, we have included a conventional factor in the numerator which ensures dual-conformal invariance, and we have used dual-momentum coordinates for which the massless external momenta are given by \mbox{$p_{\b{a}}\equivL\b{x}_{\b{a+1}}{-}\b{x}_{\b{a}}$} (with cyclic labeling understood), and with the association of $\r{\ell_i}\!\Leftrightarrow\!\r{x}_{\r{\ell_i}}$. In terms of dual-momentum coordinates, the Lorentz invariants are defined by $\x{\b{a}}{\b{b}}\equivR(\b{x}_{\b{a}}{-}\b{x}_{\b{b}})^2$.

What should the normalization of this integral be so as to make the representation of field theory amplitudes maximally transparent? One answer comes from the fact that the full-dimensional compact contour integral of the amplitude in planar sYM---which `encircles' the seven poles corresponding to the propagators of (\ref{scalar_double_box_integrand_general}) and then uses one of the fundamental cycles of the elliptic curve---is Yangian invariant \cite{Bourjaily:2020hjv}. Let us briefly review this story here, but in the general case---where the momenta flowing into the corners of the box are arbitrary.

\subsection[Elliptic Leading Singularities (and Other On-Shell Functions)]{\mbox{Elliptic Leading Singularities (\& Other On-Shell Functions)}}\label{subsec:elliptic_double_box_cut_of_sym}\vspace{0pt}

Let us start with the sub-leading singularity associated with a contour in field theory with the topology of a double-box integral as shown in (\ref{scalar_double_box_integrand_general}). That is, we'd like to define the sub-leading singularity associated with\footnote{As described in \cite{Bourjaily:2015jna}, it suffices for us to consider only MHV amplitudes at the vertices---as general amplitudes can then be generated by multiplication by the corresponding on-shell amplitudes.}
\eq{\mathfrak{db}_{\pm}(\r{\alpha})\equivR\fig{-36pt}{1.25}{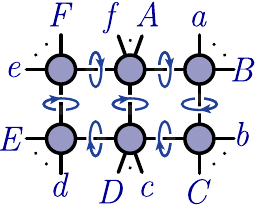}\,,\label{double_box_subleading_ls}}
where $\r{\alpha}$ represents the single on-shell degree of freedom of the sub-leading singularity, and $\pm$ denotes which of the two branches of the seven-cut equations is chosen. This could easily be represented in the Grassmannian according to \cite{ArkaniHamed:2012nw,Bourjaily:2012gy}, but we prefer to disambiguate its structure by writing it explicitly. To do this, we express $\mathfrak{db}_{\pm}(\r{\alpha})$ as simple-pole-enclosing contour integral over the co-dimension-six `kissing triangle' function, for which the non-vanishing term involves three $R$-invariants,
 \eq{\frac{\dbar{\color{hred}\alpha}\,\dbar{\color{hred}\beta}}{{\color{hred}\alpha}\,{\color{hred}\beta}}R\big[\b{1},{\color{hred}\hat{a}},{\color{hred}\hat{c}},{\color{hred}\hat{d}},{\color{hred}\hat{f}}\big]R\big[{\color{hred}\hat{a}},{\color{hblue}B},{\color{hblue}b},{\color{hblue}C},{\color{hblue}c}\big]R\big[{\color{hred}\hat{d}},{\color{hblue}E},{\color{hblue}e},{\color{hblue}F},{\color{hblue}f}\big],\label{kissing_triangle_formula_general}}
 which is a special case of the general expression given in Appendix A and B of \cite{Bourjaily:2015jna}, written here in terms of $R$-invariants with (shifted) momentum super-twistor \cite{Hodges:2009hk} arguments defined as 
 \eq{\begin{array}{l@{$\hspace{20pt}$}l}{\color{hred}\hat{a}}({\color{hred}\alpha})\,\equivR{\color{hblue}\mathcal{Z}_a}+{\color{hred}\alpha}{\color{hblue}\mathcal{Z}_A}&{\color{hred}\hat{d}}({\color{hred}\beta})\,\equivR{\color{hblue}\mathcal{Z}_d}+{\color{hred}\beta}\,{\color{hblue}\mathcal{Z}_D}\\
{\color{hred}\hat{c}}({\color{hred}\alpha})\,\equivR\big({\color{hblue}C}\,{\color{hblue}c}\big)\tncap\big({\color{hred}\hat{a}}\,{\color{hblue}B}\,{\color{hblue}b}\big)&{\color{hred}\hat{f}}({\color{hred}\beta})\,\equivR\big({\color{hblue}F}\,{\color{hblue}f}\big)\tncap\big({\color{hred}\hat{d}}\,{\color{hblue}E}\,{\color{hblue}e}\big)\,.\end{array}}
To compute the double-box sub-leading singularity, we consider a `residue' contour encircling the pole at $\ab{\r{\hat{a}\,\hat{c}\,\hat{d}\,\hat{f}}}=0$, which is a quadratic condition involving both ${\color{hred}\alpha}$ and ${\color{hred}\beta}$. Without any loss of generality, we are free to solve this condition for ${\color{hred}\beta}$; the (two) branches of solutions to $\ab{\r{\hat{a}\,\hat{c}\,\hat{d}\,\hat{f}}}=0$ are then given by 
  \eq{\hspace{-7.5pt}\r{\beta}^*_{\pm}\!\equivR\frac{\ab{{\color{hred}\hat{a}}\,{\color{hred}\hat{c}}\,{\color{hblue}d}\,\big({\color{hblue}D}\,{\color{hblue}E}\,\b{e}\big)\tncap\big({\color{hblue}F}\,{\color{hblue}f}\big)}{+}\ab{{\color{hred}\hat{a}}\,{\color{hred}\hat{c}}\,{\color{hblue}D}\,\big({\color{hblue}d}\,{\color{hblue}E}\,{\color{hblue}e}\big)\tncap\big({\color{hblue}F}\,{\color{hblue}f}\big)}\pm y(\hspace{-0.pt}\r{\alpha}\hspace{-0.pt})/c_y}{2\ab{{\color{hred}\hat{a}}\,{\color{hred}\hat{c}}\,\big({\color{hblue}D}\,{\color{hblue}E}\,{\color{hblue}e}\big)\tncap\big({\color{hblue}F}\,{\color{hblue}f}\big)\,{\color{hblue}D}}}\,,}
 where $y^2({\color{hred}\alpha})$ is a quartic polynomial defined as
  \eq{\begin{split}\hspace{-10pt}\frac{1}{c_y^2}\,y^2(\r{\alpha})\equivR\!&\Big(\!\ab{{\color{hred}\hat{a}}\,{\color{hred}\hat{c}}\,{\color{hblue}d}\,\big({\color{hblue}D}\,{\color{hblue}E}\,\b{e}\big)\tncap\big({\color{hblue}F}\,{\color{hblue}f}\big)}{+}\ab{{\color{hred}\hat{a}}\,{\color{hred}\hat{c}}\,{\color{hblue}D}\,\big({\color{hblue}d}\,{\color{hblue}E}\,\b{e}\big)\tncap\big({\color{hblue}F}\,{\color{hblue}f}\big)}\!\Big)^{\!2}\\[-5pt]
&{-}4\,\ab{{\color{hred}\hat{a}}\,{\color{hred}\hat{c}}\,{\color{hblue}D}\,\big({\color{hblue}d}\,{\color{hblue}E}\,\b{e}\big)\tncap\big({\color{hblue}F}\,{\color{hblue}f}\big)}\ab{{\color{hred}\hat{a}}\,{\color{hred}\hat{c}}\,{\color{hblue}d}\,\big({\color{hblue}D}\,{\color{hblue}E}\,\b{e}\big)\tncap\big({\color{hblue}F}\,{\color{hblue}f}\big)},\label{defn_of_y2}
\end{split}}
 and the factor of $1/c_y^2$ is included to render the quartic $y^2({\color{hred}\alpha})$ monic \emph{by construction}. The residue of the kissing-triangle hexa-cut function at ${\color{hred}\beta}\mapsto{\color{hred}\beta}_{\pm}^*$ follows by noting that 
\eq{\oint_{\ab{\r{\hat{a}\,\hat{c}\,\hat{d}\,\hat{f}}}=0}\left(\frac{\dbar\,\r{\beta}}{\ab{\r{\hat{a}\,\hat{c}\,\hat{d}\,\hat{f}}}}\right)=\pm\frac{i c_y}{\,y(\r{\alpha})}\,.}
Notice that the $\pm $ sign on the left hand side reflects the choice of the branch of the cut equations, and $y(\r{\alpha})$ is a square root of the quartic (\ref{defn_of_y2}). 

Since there are two different solutions to $\ab{\r{\hat{a}\,\hat{c}\,\hat{d}\,\hat{f}}}=0$, it makes sense to define `the' double-box sub-leading singularity $\mathfrak{db}(\r\alpha)$ as the \emph{difference} of the kissing-triangle on the two contours. We therefore define 
\eq{\Big[\mathfrak{db}_{+}(\r{\alpha})-\mathfrak{db}_{-}(\r{\alpha})\Big]\!\equivL\mathfrak{db}(\r\alpha)\equivL\frac{\dbar \r\alpha}{y(\r\alpha)}\hat{\mathfrak{db}}(\r{\alpha})\,,}
where in the second equality we have defined a useful `hatted' double-box function where the inverse of the square root of the quartic has been factored out. 

Let us now return to (\ref{double_box_subleading_ls}). The double-box on-shell function has various factorization channels corresponding to the six amplitudes appearing in its definition. Each of these channels corresponds to a simple pole of $\mathfrak{db}(\r\alpha)$ at, say, ${\color{hred}\alpha}=a_i\in\mathbb{C}$ on which we may define an additional contour about a $d\log$ pole. According to prescriptive unitarity, each such residue is topologically equivalent to a `pentabox' leading singularity, which we denote by $\mathfrak{pb}_i$. We can, therefore, expand our next-to-leading singularity in a basis of those factorizations into pentaboxes according to
\eq{\mathfrak{db}(\r{\alpha})\equivL\frac{\dbar\r{\alpha}}{y(\r{\alpha})}\mathfrak{db}_0+\frac{\dbar{\r{\alpha}}}{y(\r{\alpha})}\sum_{i}\frac{y(a_i)}{(\r{\alpha}{-}a_i)}\mathfrak{pb}_i,\label{analytic_formula_for_db_form}}
where the ${\color{hred}\alpha}$-independent coefficient of the basis element without any extra simple poles, denoted by $\mathfrak{db}_0$, is defined as:
\eq{\mathfrak{db}_0\equivR \hat{\mathfrak{db}}(\r{\alpha})-\sum_{i}\frac{y(a_i)}{(\r{\alpha}{-}a_i)}\mathfrak{pb}_i.\label{bare_piece_general_form}}
While it is not manifest in (\ref{bare_piece_general_form}), the coefficient $\mathfrak{db}_0$ is, by construction, independent of $\r{\alpha}$. This trivial---yet important---point follows directly from Liouville's theorem of elementary complex analysis: as we have removed \emph{every} pole, including the pole at $\r{\alpha}=\infty$, the `function' $\mathfrak{db}_0$ is entire on the Riemann sphere, that is, it must be a \emph{constant}. At the risk of being overly illustrative, let us emphasize that this implies that we may equivalently write
\eq{\mathfrak{db}_0\equivR \hat{\mathfrak{db}}(\r{\alpha^*})-\sum_{i}\frac{y(a_i)}{(\r{\alpha^*}{-}a_i)}\mathfrak{pb}_i,\label{bare_piece_final_form}}
where $\r{\alpha^*}$ is an arbitrarily chosen point. 

The recent work \cite{Bourjaily:2020hjv} defines an `elliptic' leading singularity by integrating the double-box seven-cut differential form $\mathfrak{db}(\r\alpha)$ over either the $a$ or $b$ `cycle' of the elliptic curve defined by $y^2(\r{\alpha})$, 
\eq{\mathfrak{e}_{a,b}\equivR\oint_{\Omega_{a,b}}\!\!\mathfrak{db}(\r{\alpha})=\oint_{\Omega_{a,b}}\!\frac{\dbar\r{\alpha}}{y(\r\alpha)}\,\hat{\mathfrak{db}}(\r{\alpha})=\pm2\oint_{\Omega_{a,b}}\!\!\mathfrak{db}_{\pm}(\r\alpha)\,.\label{elliptic_ls_v0}}
To perform the integral over the $a$-cycle, (or any cycle for that matter) it is quite useful to factorize the monic quartic polynomial defined in (\ref{defn_of_y2}) in terms of its roots,
\eq{y^2(\r\alpha)\equivL(\r\alpha-r_1)(\r\alpha-r_2)(\r\alpha-r_3)(\r\alpha-r_4)\,,}
where, for positive kinematics \cite{Arkani-Hamed:2013jha}, the roots form two complex conjugate pairs $\{r_1,r_2\}$ and $\{r_3,r_4\}$, and are ordered so that $\mathfrak{Re}(r_1)\!>\!\mathfrak{Re}(r_3)$ and $\mathfrak{Im}(r_{1,3})\!>\!0$. With this particular ordering in mind, the $a,b$-cycle integrals (\ref{elliptic_ls_v0}) can be written in terms of elliptic integrals involving the dual-conformal invariant cross-ratio
\eq{\phi\equivR\frac{(r_2{-}r_1)(r_3{-}r_4)}{(r_2{-}r_3)(r_1{-}r_4)}\equivL\, \frac{r_{21}r_{34}}{r_{23}r_{14}}\,,}
which, in our conventions, is always within the interval $[0,1]$ for positive kinematics. Choosing the branch cuts to connect each complex conjugate pair of roots, we now define the $a$ cycle contour, $\Omega_a$, to enclose the cut between $r_{1,2}$. To compute the full $a$-cycle integral we use the two formulae
\eq{\oint_{\Omega_a}\,\,\frac{\dbar\r\alpha}{y(\r\alpha)}=\frac{2\,i}{\pi\sqrt{r_{32}r_{41}}}K[\phi]\,,\label{first_kind_period_a}}
and 
\eq{\fwbox{0pt}{\hspace{-10pt}\oint_{\Omega_a}\!\!\!\frac{\dbar\r\alpha\,y(a_i)}{(\r\alpha{-}a_i)y(\r\alpha)}\!=\!\frac{2\,i}{\pi\sqrt{r_{32}r_{41}}}\frac{y(a_i)}{(r_4{-}a_i)}\Bigg(\!\!K[\phi]\!+\!\!\frac{r_{42}}{(r_2{-}a_i)}\Pi\left[\frac{(r_4{-}a_i)r_{21}}{(r_2{-}a_i)r_{41}},\phi\right]\!\!\Bigg),}\label{third_kind_period_a}}
where our definitions of the complete elliptic integrals of the first and third kinds, $K[\phi]$ and $\Pi[q,\phi]$, respectively, are in agreement with \textsc{Mathematica}'s. We have provided a different---and perhaps more easily generalizable---representation of these elliptic period integrals in terms of Lauricella functions in \mbox{appendix \ref{hypergeomtric_appendix}}. The attentive reader will notice that these two integral formulae differ from the results quoted in \cite{Bourjaily:2020hjv}; this discrepancy is a consequence of our use of $\dbar=d/(2\pi)$ throughout this work. Following the discussion in \cite{Bourjaily:2020hjv}, observe that in the definition (\ref{analytic_formula_for_db_form}) of the double-box one-form, the coefficient of the pentabox $\mathfrak{pb}_i$ is the same as in (\ref{bare_piece_final_form}), but for the appearance of $\r{\alpha^*}$ in place of $\r{\alpha}$. Upon performing the integration, if we choose $\r{\alpha}\rightarrow r_4$ the terms involving $K[\phi]\mathfrak{pb}_i$ thus cancel, and we are left with
\eq{\fwbox{0pt}{\hspace{-30pt}\mathfrak{e}_a=\frac{2 i}{\pi\sqrt{r_{32}r_{41}}}\Bigg(K[\phi]\,\,\hat{\mathfrak{db}}(\r{\alpha^*}\!\!\to r_4)+\sum_{i}\frac{y(a_i)r_{42}}{(r_2{-}a_i)(r_4{-}a_i)}\Pi\left[\frac{(r_4{-}a_i)r_{21}}{(r_2{-}a_i)r_{41}},\phi\right]\mathfrak{pb}_i\!\!\Bigg)\!.} \label{simple_form_of_ls}}

If instead we consider the $b$-cycle, say the one encircling a branch cut which connect two roots with different real part (e.g. $r_1$ and $r_3$), we have
\eq{\mathfrak{e}_b=\frac{2 i}{\pi\sqrt{r_{23}r_{41}}}\Bigg(K[1{-}\phi]\,\,\hat{\mathfrak{db}}(\r{\alpha^*}\!\!\to r_4)+\sum_{i}\frac{y(a_i)r_{43}}{(r_3{-}a_i)(r_4{-}a_i)}\Pi\left[\frac{(r_4{-}a_i)r_{31}}{(r_3{-}a_i)r_{41}},1{-}\phi\right]\mathfrak{pb}_i\!\!\Bigg),\label{simple_form_of_ls_contour_b}}
which is the same with the first expression after the exchange $r_2\leftrightarrow r_3$. As discussed at greater length in \cite{Bourjaily:2020hjv}, both expressions (\ref{simple_form_of_ls}) and (\ref{simple_form_of_ls_contour_b}) are non-trivially Yangian invariant, as may be verified by direct computation using, for example, the level one generators written in momentum super-twistor space. 
 
\newpage\vspace{0pt}%
\section{New Prescriptive Representations: Two Approaches}\label{sec:new_prescriptive_integrands}\vspace{0pt}

We shall now argue that the results of the previous section suggest two natural prescriptions for the normalization of the double-box integrand which avoid entirely the arbitrary choices of the original prescriptive unitarity program.

\vspace{0pt}%
\subsection{\emph{Homological} Diagonalization---with Respect to \emph{Contours}}\label{subsec:homological_diagonalization}\vspace{0pt}
One natural choice for the normalization of the double-box integrand $\mathbf{n}_{\text{db}}$ would be analogous to the choice made in the introduction for scalar $\phi^4$ theory. That is, we may choose to normalize the integrand so that it \emph{integrates} to 1 on the contour associated with, say, the $a$-cycle of the elliptic curve; since the loop-momentum-dependent part of the integrand is proportional to $1/y(\r{\alpha})$, this fixes its normalization, using (\ref{first_kind_period_a}), to be
\eq{\label{hom_norm}\mathbf{n}_{\text{db}}\equivR\frac{\pi}{2}\frac{\sqrt{(r_3{-}r_2)(r_4{-}r_1)}}{c_y\,K[\phi]}\,. }
This choice of normalization ensures that we match the elliptic leading singularity in field theory, $\mathfrak{e}_a$, \emph{manifestly} with the double-box integrand directly
\eq{\oint_{\Omega_a}\oint_{\substack{\{|\x{\r{\ell_i}}{\b{a}}|=\epsilon\}\\|\x{\r{\ell_1}}{\r{\ell_2}}|=\epsilon}}\mathcal{A}=\mathfrak{e}_a\oint_{\Omega_a}\oint_{\substack{\{|\x{\r{\ell_i}}{\b{a}}|=\epsilon\}\\|\x{\r{\ell_1}}{\r{\ell_2}}|=\epsilon}}\mathcal{I}_{\text{db}}=\mathfrak{e}_a\,.}

Notice that this representation, however, does not match the $b$-cycle leading singularity of the amplitude manifestly at all! Although we have normalized the double-boxes so that the $a$-cycle contours of field theory are manifest, the result on the $b$-cycle is, rather, 
\begin{align}&\mathfrak{e}_a\oint_{\Omega_b}\oint_{\substack{\{|\x{\r{\ell_i}}{\b{a}}|=\epsilon\}\\|\x{\r{\ell_1}}{\r{\ell_2}}|=\epsilon}}\mathcal{I}_{\text{db}}=-i\,\mathfrak{e}_a\,\frac{K[1{-}\phi]}{K[\phi]}\label{double_box_times_ea_on_b_cycle}\\
&=\!\frac{2}{\pi\sqrt{r_{23}r_{41}}}\Bigg(\!\!K[1{-}\phi]\,\,\hat{\mathfrak{db}}(\r{\alpha^*}\!\!\to r_4){+}\sum_{i}\!\frac{y(a_i)r_{42}}{(r_2{-}a_i)(r_4{-}a_i)}\frac{K[1{-}\phi]}{K[\phi]}\Pi\left[\frac{(r_4{-}a_i)r_{21}}{(r_2{-}a_i)r_{41}},\phi\right]\!\mathfrak{pb}_i\!\!\Bigg)\!\nonumber\end{align}
which is not at all equal to
\eq{\oint_{\Omega_a}\oint_{\substack{\{|\x{\r{\ell_i}}{\b{a}}|=\epsilon\}\\|\x{\r{\ell_1}}{\r{\ell_2}}|=\epsilon}}\mathcal{A}=\mathfrak{e}_b\,}
which was given in (\ref{simple_form_of_ls_contour_b}). 
This is not in fact a problem: as we will see, the pentaboxes (and double-pentagon integrands) \emph{do} have support on the $b$-cycle integrals (even after their contact terms have been fixed), and exactly give the necessary contributions to reproduce the correct $b$-cycle leading singularity in (\ref{simple_form_of_ls_contour_b}). (With hindsight, this `magic' can be seen to follow directly from the completeness of the integrand basis.)

Let us now discuss the implications of prescriptive unitarity for the contact-term rules of the pentabox and double-pentagon integrands. As always, prescriptivity requires that our integrands be \emph{diagonal} in a choice of contours; therefore, in the homological scheme, the contact terms are determined by the requirement that all other integrands in our basis \emph{vanish} identically on all elliptic $\Omega_a$-cycle contours associated with double-box sub-topologies. 

To see how this works in practice, consider a pentabox integrand which contains a double-box contact-term:
\eq{\hspace{-10pt}\begin{tikzpicture}[scale=1*\figScale,baseline=-3.05,rotate=0]
\dBoxCoords\dBoxEdges
\leg{(v1)}{90};\leg{(v2)}{45};\leg{(v3)}{-45};\leg{(v4)}{-90};\leg{(v5)}{-135};\leg{(v6)}{135};
\foreach\n in {1,...,6}{\node at (v\n) [ddot] {};};
\fill[hdot]($($(v1)!.5!(v2)$)+(90:0.2)$) circle (2pt) node[above=2pt] {\normalsize $\b{a}$};
\fill[hdot]($($(v2)!.5!(v3)$)+(0:0.2)$) circle (2pt) node[right=1pt] {\normalsize $\b{b}$};
\fill[hdot]($($(v3)!.5!(v4)$)+(-90:0.2)$) circle (2pt) node[below=2pt] {\normalsize $\b{c}$};
\fill[hdot]($($(v4)!.5!(v5)$)+(-90:0.2)$) circle (2pt) node[below=1pt] {\normalsize $\!\b{e}\,$};
\fill[hdot]($($(v5)!.5!(v6)$)+(180:0.2)$) circle (2pt) node[left=1pt] {\normalsize $\b{f}$};
\fill[hdot]($($(v6)!.5!(v1)$)+(90:0.2)$) circle (2pt) node[above=1pt] {\normalsize $\!\b{g}\,$};
\node at ($($(v2)!.5!(v3)$)+(180:0.4)$)  {{\normalsize $\r{\ell_1}$}};
\node at ($($(v1)!.5!(v4)$)+(180:0.4)$)  {{\normalsize $\r{\ell_2}$}};
\end{tikzpicture}
\hspace{0pt}\bigger{\subset}\hspace{0pt}
\begin{tikzpicture}[scale=1*\figScale,baseline=-3.05,rotate=0]
\pBoxCoords\draw[int](v2)--(v3)--(v4)--(v5)--(v6)--(v7)--(v1);\draw[int](v4)--(v7);\draw[intH](v1)--(v2);
\fill[hdot]($($(v7)!.5!(v1)$)+(108:0.2)$) circle (2pt) node[above=2pt] {\normalsize $\b{a}$};
\fill[fill=hred,circle,minimum size=0.15*\dotSize,inner sep=0]($($(v1)!.5!(v2)$)+(36:0.2)$) circle (2pt) node[right=1pt] {\normalsize $\r{\rho}$};
\fill[hdot]($($(v2)!.5!(v3)$)+(-36:0.2)$) circle (2pt) node[right=1pt] {\normalsize $\b{b}$};
\fill[hdot]($($(v3)!.5!(v4)$)+(-108:0.2)$) circle (2pt) node[below=2pt] {\normalsize $\b{c}$};
\fill[hdot]($($(v4)!.5!(v5)$)+(-90:0.2)$) circle (2pt) node[below=1pt] {\normalsize $\!\b{e}\,$};
\fill[hdot]($($(v5)!.5!(v6)$)+(180:0.2)$) circle (2pt) node[left=1pt] {\normalsize $\b{f}$};
\fill[hdot]($($(v6)!.5!(v7)$)+(90:0.2)$) circle (2pt) node[above=1pt] {\normalsize $\!\b{g}\,$};
\node at ($($(v2)$)+(180:0.75)$)  {{\normalsize $\mathbf{n}^i_{\r{\rho}}\hspace{-0.5pt}(\hspace{-0.85pt}\r{\ell_1}\hspace{-0.85pt})$}};
\node at ($($(v7)!.5!(v4)$)+(180:0.4)$)  {{\normalsize $\r{\ell_2}$}};
\leg{(v1)}{72};\leg{(v2)}{0};\leg{(v3)}{-72};\leg{(v4)}{-105};\leg{(v5)}{-135};\leg{(v6)}{135};\leg{(v7)}{105};
\end{tikzpicture}\bigger{\;\Leftrightarrow\;}\frac{\dbar^4\r{\ell_1}\,\dbar^4\r{\ell_2}\,\,\left[\mathbf{n}_0^{i}\x{\r{\ell_1}}{\b{\mathcal{Y}^i}}{+}\mathbf{n}^i_{\r{\rho}}\x{\r{\ell_1}}{\r{\rho}}{+}\ldots\right]}{\x{\r{\ell_1}}{\b{a}}\x{\r{\ell_1}}{\b{b}}\x{\r{\ell_1}}{\b{c}}\x{\r{\ell_1}}{\r{\rho}}\x{\r{\ell_1}}{{\ell_2}}\x{{\ell_2}}{{\b{e}}}\x{{\ell_2}}{{\b{f}}}\x{{\ell_2}}{{\b{g}}}}}
where $\mathbf{n}^i_{\r{\rho}}$ is the coefficient of the term in the numerator proportional to $\x{\r{\ell_1}}{\r{\rho}}$. Schematically, the full contribution of the pentabox on the $a$-cycle contour can be written as
\eq{\oint_{\Omega_a}\oint_{\substack{\{|\x{\r{\ell_i}}{\b{a}}|=\epsilon\}\\|\x{\r{\ell_1}}{\r{\ell_2}}|=\epsilon}}\begin{tikzpicture}[scale=1*\figScale,baseline=-3.05,rotate=0]
\pBoxCoords\draw[int](v2)--(v3)--(v4)--(v5)--(v6)--(v7)--(v1);\draw[int](v4)--(v7);\draw[intH](v1)--(v2);
\fill[hdot]($($(v7)!.5!(v1)$)+(108:0.2)$) circle (2pt) node[above=2pt] {\normalsize $\b{a}$};
\fill[fill=hred,circle,minimum size=0.15*\dotSize,inner sep=0]($($(v1)!.5!(v2)$)+(36:0.2)$) circle (2pt) node[right=1pt] {\normalsize $\r{\rho}$};
\fill[hdot]($($(v2)!.5!(v3)$)+(-36:0.2)$) circle (2pt) node[right=1pt] {\normalsize $\b{b}$};
\fill[hdot]($($(v3)!.5!(v4)$)+(-108:0.2)$) circle (2pt) node[below=2pt] {\normalsize $\b{c}$};
\fill[hdot]($($(v4)!.5!(v5)$)+(-90:0.2)$) circle (2pt) node[below=1pt] {\normalsize $\!\b{e}\,$};
\fill[hdot]($($(v5)!.5!(v6)$)+(180:0.2)$) circle (2pt) node[left=1pt] {\normalsize $\b{f}$};
\fill[hdot]($($(v6)!.5!(v7)$)+(90:0.2)$) circle (2pt) node[above=1pt] {\normalsize $\!\b{g}\,$};
\node at ($($(v2)$)+(180:0.75)$)  {{\normalsize $\mathbf{n}^i_{\r{\rho}}\hspace{-0.5pt}(\hspace{-0.85pt}\r{\ell_1}\hspace{-0.85pt})$}};
\node at ($($(v7)!.5!(v4)$)+(180:0.4)$)  {{\normalsize $\r{\ell_2}$}};
\leg{(v1)}{72};\leg{(v2)}{0};\leg{(v3)}{-72};\leg{(v4)}{-105};\leg{(v5)}{-135};\leg{(v6)}{135};\leg{(v7)}{105};
\end{tikzpicture}
=\oint_{\Omega_a}\dbar\r{\alpha}\,\left[\frac{y(a_{\r{\rho}})}{(\r{\alpha}{-}a_{\r\rho})y(\r{\alpha})}+\frac{\mathbf{n}^i_{\r{\rho}}}{y(\r{\alpha})}\right]\,,
\label{pentabox_cont_on_a_cycle}}
where $a_{\r{\rho}}$ corresponds to the pole where $\x{\r{\ell_1}}{\r{\rho}}\!=\!0$. Notice that the leading term follows directly from the fact that the integral is normalized to have \emph{unit} leading singularity on some pentabox contour. Diagonalization of the basis according to homology---leading singularities---fixes this contact-term coefficient, $\mathbf{n}^i_{\r{\rho}}$, of the pentabox by the criterion that (\ref{pentabox_cont_on_a_cycle}) vanishes. Namely, we must choose 
\eq{\mathbf{n}^i_{\r{\rho}}\equivR-\frac{y(a_{\r\rho})}{(r_4-a_{\r\rho})}\Bigg[1+\frac{r_{42}}{(r_2-a_{\r\rho})}\Pi\left[\frac{(r_4{-}a_i)r_{21}}{(r_2{-}a_i)r_{41}},\phi\right]/K[\phi]\Bigg]\,.}%

One important thing to note is that the pentabox integrands with these contact-terms chosen \emph{do not vanish} on the $b$-cycle contours. This is a good thing!—as the normalization we have chosen for the double-boxes makes the correctness of the amplitude integrand on the $b$-cycle extremely non-manifest in this representation. It is not hard to see, however, that when these contact terms are used, they generate on the $b$-cycle exactly the terms needed to cancel the `wrong' elliptic-$\Pi$ terms involving the pentabox leading singularities that arise from the double-box integrands in (\ref{double_box_times_ea_on_b_cycle}). Furthermore, the contact terms of the pentabox integrands also \emph{contribute} the correct pieces involving the pentabox leading singularities appearing in $\mathfrak{e}_b$ (\ref{simple_form_of_ls_contour_b}) once the corresponding contributions from the double-pentagons are included (so as to match all pentabox leading singularities).\\ 

This diagonalization strategy has several obvious advantages. For one thing, it is morally the direct realization of `prescriptive unitarity' according to a choice of leading singularities. Moreover, as emphasized in the introduction, it should have the property that all integrands defined in this way are \emph{pure}, and all coefficients are Yangian-invariant. 

Nevertheless, there are several reasons to be dissatisfied with this basis of integrands. For example, it deeply obscures the fact that the integrand is a \emph{rational} differential form in loop momenta. Choosing basis integrands whose normalization depends on the roots of quartics makes this representation fairly unwieldy in practice (at least for most computer algebra packages). Therefore, we are motivated to consider a slightly different strategy, with huge advantages in terms of (algebraic) complexity, but which abandons the desire for a pure integrand basis---and requires the use of non-Yangian-invariant coefficients.

\vspace{0pt}%
\subsection{\emph{Cohomological} Diagonalization---with Respect to \emph{Forms}}\label{subsec:cohomological_diagonalization}\vspace{0pt}
The attentive reader may already have guessed an alternative strategy for matching amplitudes in sYM---namely, according to the various differential forms that appear in the double-box sub-leading singularity $\mathfrak{db}(\r\alpha)$ in (\ref{analytic_formula_for_db_form}):
\eq{\mathfrak{db}(\r{\alpha})\equivL\frac{\dbar\r{\alpha}}{y(\r{\alpha})}\mathfrak{db}_0+\frac{\dbar{\r{\alpha}}}{y(\r{\alpha})}\sum_{i}\frac{y(a_i)}{(\r{\alpha}{-}a_i)}\mathfrak{pb}_i\,.\label{analytic_formula_for_db_form_2}}
Considering the fact that the co-dimension seven contour of the scalar double-box integrand (\ref{scalar_double_box_integrand_general}) which encircles its seven propagators results in 
\eq{\oint_{\substack{\{|\x{\r{\ell_i}}{\b{a}}|=\epsilon\}\\|\x{\r{\ell_1}}{\r{\ell_2}}|=\epsilon}}\dbar^4\r{\ell_1}\,\dbar^4\r{\ell_2}\frac{\mathbf{n}_{\text{db}}\,\x{\b{a}}{\b{c}}\x{\b{d}}{\b{f}}\x{\b{b}}{\b{e}}}{\x{\r{\ell_1}}{\b{a}}\x{\r{\ell_1}}{\b{b}}\x{\r{\ell_1}}{\b{c}}\x{\r{\ell_1}}{\r{\ell_2}}\x{\r{\ell_2}}{\b{d}}\x{\r{\ell_2}}{\b{e}}\x{\r{\ell_2}}{\b{f}}}=\mathbf{n}_{\text{db}}(-i\,c_y)\frac{\dbar\r\alpha}{y(\r\alpha)}\,,}
it would be natural to choose $\mathbf{n}_{\text{db}}$ to be $+i$, and to match its coefficient in the representation of field theory amplitudes to be $c_y\mathbf{db}_0$. (With this factor of $c_y$---implicit in the definition (\ref{analytic_formula_for_db_form_2})---this on-shell function is little-group neutral (if not Yangian-invariant).) Moreover, it is easy to see that this combination of integrand and coefficient automatically matches the \emph{leading} term in $\mathfrak{e}_b$ in (\ref{simple_form_of_ls_contour_b}). The only terms missing from both leading singularities are those involving the pentabox leading singularities $\mathfrak{pb}_i$. 

These not-yet-matched pieces of $\mathfrak{e}_{a,b}$ can easily be seen to arise from the pentabox and double-pentagon contributions. Consider again how the contact terms appear in the pentabox integrands' contributions to the $a$-cycle, say:
\eq{\oint_{\Omega_a}\oint_{\substack{\{|\x{\r{\ell_i}}{\b{a}}|=\epsilon\}\\|\x{\r{\ell_1}}{\r{\ell_2}}|=\epsilon}}\begin{tikzpicture}[scale=1*\figScale,baseline=-3.05,rotate=0]
\pBoxCoords\draw[int](v2)--(v3)--(v4)--(v5)--(v6)--(v7)--(v1);\draw[int](v4)--(v7);\draw[intH](v1)--(v2);
\fill[hdot]($($(v7)!.5!(v1)$)+(108:0.2)$) circle (2pt) node[above=2pt] {\normalsize $\b{a}$};
\fill[fill=hred,circle,minimum size=0.15*\dotSize,inner sep=0]($($(v1)!.5!(v2)$)+(36:0.2)$) circle (2pt) node[right=1pt] {\normalsize $\r{\rho}$};
\fill[hdot]($($(v2)!.5!(v3)$)+(-36:0.2)$) circle (2pt) node[right=1pt] {\normalsize $\b{b}$};
\fill[hdot]($($(v3)!.5!(v4)$)+(-108:0.2)$) circle (2pt) node[below=2pt] {\normalsize $\b{c}$};
\fill[hdot]($($(v4)!.5!(v5)$)+(-90:0.2)$) circle (2pt) node[below=1pt] {\normalsize $\!\b{e}\,$};
\fill[hdot]($($(v5)!.5!(v6)$)+(180:0.2)$) circle (2pt) node[left=1pt] {\normalsize $\b{f}$};
\fill[hdot]($($(v6)!.5!(v7)$)+(90:0.2)$) circle (2pt) node[above=1pt] {\normalsize $\!\b{g}\,$};
\node at ($($(v2)$)+(180:0.75)$)  {{\normalsize $\mathbf{n}^i_{\r{\rho}}\hspace{-0.5pt}(\hspace{-0.85pt}\r{\ell_1}\hspace{-0.85pt})$}};
\node at ($($(v7)!.5!(v4)$)+(180:0.4)$)  {{\normalsize $\r{\ell_2}$}};
\leg{(v1)}{72};\leg{(v2)}{0};\leg{(v3)}{-72};\leg{(v4)}{-105};\leg{(v5)}{-135};\leg{(v6)}{135};\leg{(v7)}{105};
\end{tikzpicture}
=\oint_{\Omega_a}\dbar\r{\alpha}\,\left[\frac{y(a_{\r{\rho}})}{(\r{\alpha}{-}a_{\r\rho})y(\r{\alpha})}+\frac{\mathbf{n}^i_{\r{\rho}}}{y(\r{\alpha})}\right]\,.\label{pentabox_contact_term_second_time}}
Taking the contact-term numerator $\mathbf{n}^i_{\r{\rho}}\!\mapsto\!0$ will leave a non-vanishing contribution from the first term in (\ref{pentabox_contact_term_second_time}) on both cycles. It is easy to see that the missing differential form (which results in the combination of the elliptic functions $K$ and $\Pi$) in (\ref{third_kind_period_a}) is exactly that which appears in the pentabox integrands already---and the coefficients of these integrands are precisely the $\mathfrak{pb}_i$ needed to reproduce the `missing' pieces in the elliptic leading singularities $\mathfrak{e}_{a,b}$.

This definition of the double-box and the corresponding rule for the contact terms of the pentabox and double-pentagon integrands results in an extremely simple prescription for the integrand. Moreover, it is morally equivalent to a choice of diagonalization with respect to the various (local) \emph{differential-forms} in loop momentum space. As such, we call such a prescription a \emph{cohomological} choice for our basis. 

Despite the obvious advantages, we have checked that the coefficient of the double-box normalized in this way (namely, $c_y\mathfrak{db}_0$) is \emph{not Yangian-invariant}. Moreover, the double-box integrands are not pure. We strongly suspect that the pentabox and double-pentagon integrands (that contain elliptic contact-term components) are similarly non-pure. Thus, despite the algebraic and conceptual simplicity of the cohomological approach just described, we suspect that the \emph{homological} prescriptive representation will ultimately prove the superior one for integration.

\vspace{2pt}%
\subsection{Consistency Checks for Amplitude Integrands}\label{subsec:checks}\vspace{0pt}
As already mentioned in section~\ref{sec:prescriptive_unitarity_review}, the pentabox integrands lack the requisite number of degrees of freedom to match all pentabox cuts of field theory term-by-term. However, the expressions for the elliptic leading singularities $\mathfrak{e}_{a,b}$ involve them separately---and arising at poles in \emph{different} locations in the $\r\alpha$-plane. Thus, our description above regarding the pentaboxes' roles in these leading singularities does not alone ensure that we have matched everything in field theory. The missing ingredient, of course, are the kissing-box leading singularities times double-pentagons. These terms \emph{when combined with those of the pentaboxes} ensure that all the pentabox leading singularities \emph{do} get matched correctly, and individually. Thus, we have been somewhat schematic in our analysis above, relying on the fact that these parts of the amplitude integrands are guaranteed to work correctly once taken in combination. 

While ensured to work, we have checked this completely in the case of the 10-particle N$^3$MHV amplitude. Specifically, we have checked that both diagonalization procedures described above---the homological and the cohomological---result in integrand representations that exactly match the results of BCFW recursion, say. Thus, we are confident this procedure is free of any over-looked subtleties.

\vspace{0pt}%
\subsection{Smooth Degenerations}\label{subsec:degenerations}\vspace{0pt}

The reader may have considered our preference for normalizing integrands with respect to the $a$-cycle $\Omega_a$ rather arbitrary. It was not entirely so. As discussed in \cite{Bourjaily:2020hjv}, the $a$-cycle integral of the elliptic curve smoothly degenerates to $1$ in all relevant cases (or kinematic limits). Thus, our analysis above, if applied to the case of double-box integrands that \emph{do} support polylogarithmic contour integrals, reduces naturally to ordinary polylogarithmic ones. (Recall that a unit-residue integrand $d^{4L}\vec{\r\ell}$ is equivalent to a unit-\emph{contour} integrand for $\dbar^{4L}\vec{\r\ell}$.) Thus, we can apply the results of this work to truly general double-boxes, and thereby construct pure-integrand bases that happen to be `$d\log$' whenever the double-boxes' elliptic curves turn out to be degenerate. 

\vspace{0pt}%
\section{Generalizations: More Loops and Calabi-Yau Manifolds}\label{sec:generalizations}\vspace{0pt}

Beyond two loops (and for non-planar theories at two loops), non-polylogarithmic structures beyond elliptic integrals abound \cite{Bourjaily:2017bsb,Bourjaily:2018ycu,Bourjaily:2018yfy,Bourjaily:2019hmc,Vergu:2020uur}. Examples of scalar integrals with such structures include the three-loop traintrack and wheel integrals,\\[-10pt]
\eq{\left\{\begin{tikzpicture}[scale=0.8*\figScale,baseline=-3.05,rotate=0]
\traintrackCoords\draw[int](v1)--(v2)--(v3)--(v4)--(v5)--(v6)--(v7)--(v8)--(v1);\draw[int](v1)--(v6);\draw[int](v2)--(v5);
\leg{(v1)}{90};\leg{(v2)}{90};\leg{(v3)}{45};\leg{(v4)}{-45};\leg{(v5)}{-90};\leg{(v6)}{-90};\leg{(v7)}{-135};\leg{(v8)}{135};
\end{tikzpicture}\,,\,\begin{tikzpicture}[scale=0.8*\figScale,baseline=-3.05,rotate=0]
\wheelCoords\draw[int](v0)--(v1);\draw[int](v0)--(v3);\draw[int](v0)--(v5);\draw[int](v1)--(v2)--(v3)--(v4)--(v5)--(v6)--(v1);
\leg{(v1)}{0};\leg{(v2)}{60};\leg{(v3)}{120};\leg{(v4)}{180};\leg{(v5)}{240};\leg{(v6)}{300};\node[ddot] at (v0) {};
\end{tikzpicture}\right\},}%
the maximal cuts (sub-leading singularities encircling all propagators) of which are known to involve Calabi-Yau 2- and 3-folds, respectively. 

To see how the corresponding non-polylogarithmic leading singularities can be incorporated analogously to what we have described for the elliptic case, let us briefly outline the structure expected for the traintrack contribution. 

On the maximal cut surface encircling all 10 propagators of the traintrack, the sub${}^2$-leading singularity should take the form:
\eq{\mathfrak{tr}(\r{\alpha},\r{\beta})\equivL\frac{\dbar\r{\alpha}\,\dbar\r{\beta}}{y(\r{\alpha},\r{\beta})}\hat{\mathfrak{tr}}(\r{\alpha},\r{\beta})\,,\label{traintrack_10_cut}}
where $y^2(\r{\alpha},\r{\beta})$ is an irreducible quartic in both variables $\{\r\alpha,\r\beta\}$ simultaneously \cite{Bourjaily:2018ycu}. As with the double-box sub-leading singularity (\ref{double_box_subleading_ls}), the traintrack in sYM will have multiple co-dimension one, simple poles around which there are elliptic sub-leading singularities. We may express this by decomposing (\ref{traintrack_10_cut}) according to 
\eq{\hat{\mathfrak{tr}}(\r{\alpha},\r{\beta})\equivL\hat{\mathfrak{tr}}_0+\sum_{a_i}\frac{y^2(a_i,\r\beta)}{(\r\alpha-a_i)y(a_i,\r\beta)}\mathfrak{el}_{a_i}(\r{\beta})+\sum_{b_i}\frac{y^2(\r\alpha,b_i)}{(\r\beta-b_i)y(\r\alpha,b_i)}\mathfrak{el}_{b_i}(\r{\alpha})\,,}
where the terms in the sum $\mathfrak{el}_{a_i}$ and $\mathfrak{el}_{b_i}$ are elliptic sub-leading singularities arising from single-pole factorizations of the original traintrack. These in turn can be further decomposed analogously to (\ref{analytic_formula_for_db_form}), resulting in an expression of the form 
\eq{\begin{split}\hat{\mathfrak{tr}}(\r{\alpha},\r{\beta})\equivL&\hat{\mathfrak{tr}}_0+\sum_{a_i}\frac{y^2(a_i,\r\beta)}{(\r\alpha-a_i)y(a_i,\r\beta)}\hat{\mathfrak{el}}_{a_i}+\sum_{b_i}\frac{y^2(\r\alpha,b_i)}{(\r\beta-b_i)y(\r\alpha,b_i)}\hat{\mathfrak{el}}_{b_i}\\&+\sum_{a_i,b_j}\frac{y(a_i,b_j)}{(\r\alpha-a_i)(\r\beta-b_j)}\mathfrak{pl}_{a_i,b_j},\end{split}\label{traintrack_ls_decomposition}}
where $\mathfrak{pl}_{a_i,b_j}$ are the `penta-ladder' polylogarithmic leading singularities associated with simultaneous factorizations in two different loops.

The number and detailed form of terms appearing in this decomposition will depend on the number of factorization channels of the traintrack (which depends on multiplicity), but the basic  structure is clear: (\ref{traintrack_ls_decomposition}) is nothing but a decomposition of the sub${}^2$-leading singularity (\ref{traintrack_10_cut}) into a basis of differential forms involving one or two simple poles, respectively---with superfunction coefficients. (For the scalar traintrack contribution to the three-loop 12-particle amplitude in scalar $\phi^4$ theory, only the leading term $\hat{\mathfrak{tr}}_0$ is required---which happens to be $(\mi1)$---as there are no factorization channels to the participating $\phi^4$ tree-amplitudes.)

The generalization of this analysis to the three-loop wheel integral---which involves a Calabi-Yau three-fold surface---is relatively straightforward, resulting in a decomposition of the sub${}^3$-leading singularity into a top-level, irreducible volume-form times some `CY${}_3$' leading-singularity, three separate sums of $K3$-volume forms with single simple poles times `$K3$' leading singularities; three, double-nested sums of elliptic integrals with two simple poles times `elliptic' leading singularities; and, finally, a triple-sum of terms with simple poles times polylogarithmic leading singularities. 

It is worth mentioning that, unlike the three-loop traintrack which is \emph{known} to have support in sYM (as argued in \cite{Bourjaily:2018ycu}), the three-loop wheel is not any single component of an amplitude in planar sYM; as such, it is \emph{possible} that the CY${}_3$ leading singularity vanishes. Thus, this makes its evaluation an important open challenge---left for future work. (While we are unaware of closed analytic formulae for the period integrals that would be required for such a check---analogous to those in (\ref{first_kind_period_a}) and (\ref{third_kind_period_a}) for the elliptic case---we are relatively optimistic that numerical integration will work for these low-dimensional cases.)

The implications of these higher-dimensional Calabi-Yau leading singularities for prescriptive unitarity should be clear. In particular, we suspect that if the three-loop basis of planar integrands outlined in \cite{Bourjaily:2017wjl} were diagonalized \emph{homologically}, the result would be a complete representation of amplitudes involving term-wise `pure' integrals times Yangian-invariants; and the  \emph{cohomological} diagonalization of this basis into separate forms should be extremely straightforward to implement from the decompositions as in (\ref{traintrack_ls_decomposition}).

\newpage\vspace{0pt}%
\section{Conclusions and Future Directions}\label{sec:conclusions}\vspace{0pt}
In this work we have made use of the new, broadened definition of leading singularities (beyond the polylogarithmic case) to derive two new prescriptive representations of two-loop scattering amplitude integrands in planar sYM. This analysis was illustrative of a more general strategy, with applications well beyond the planar limit and to theories with less or no supersymmetry. In many ways, our results directly reflect the primary goals of \emph{prescriptive} unitarity: constructing loop-integrand bases that are diagonal in a spanning set of \emph{contours}. For scattering amplitudes free of non-polylogarithmic structures, this strategy directly reproduces the $d\log$ differential forms of the traditional approach; but as we have seen, the generalization beyond polylogarithms is extremely natural. We have also motivated a different strategy: how to construct a prescriptive integrand basis diagonal with respect to \emph{cohomology}; the resulting basis may not be pure and the coefficients required may not be Yangian-invariant, but the ultimate representation of loop integrands is dramatically simpler from an algebraic point of view. 

We have described how this story illustrates a broader one---with applications well beyond the case of elliptic leading singularities in planar theories at two loops. It would be extremely interesting to apply these lessons more widely to generate (purportedly) `pure' master integrals for applications beyond the planar limit, and to theories without supersymmetry. Although we have not proven the `purity' of this broader class of integrals, and although the strategies and techniques required to efficiently exploit the differential structure of pure integrals are still being developed (even in the elliptic case---but see e.g. \cite{Adams:2018yqc,Weinzierl:2019pfw,Walden:2020odh}), we strongly suspect that the prescriptive bases we have constructed will prove computationally valuable as master integrals for diverse applications. For example, constructing such bases for massive theories now appears straightforward, and undoubtedly has more immediate applications for real physical applications (see e.g. \cite{Chaubey:2019lum,Weinzierl:2019pfw,Weinzierl:2020kyq,Weinzierl:2020gda,Walden:2020odh}). But we leave such analyses to future work.

\vspace{\fill}\vspace{-4pt}
\section*{Acknowledgements}%
\vspace{-4pt}
\noindent The authors gratefully acknowledge fruitful contributions from Marcus Spradlin during the early stages of this work, and for fruitful conversations with Nima Arkani-Hamed, Song He, Enrico Herrmann, Jaroslav Trnka, and Cristian Vergu. 
This work was performed in part at the Aspen Center for Physics, which is supported by National Science Foundation grant PHY-1607611, and the Harvard Center of Mathematical Sciences and Applications. 
This project has been supported by an ERC Starting Grant \mbox{(No.\ 757978)}, a grant from the Villum Fonden \mbox{(No.\ 15369)}, by a grant from the Simons Foundation (341344, LA) (JLB).

\newpage

\appendix
\section{Hypergeometric Representations of the Elliptic Integrals}\label{hypergeomtric_appendix}

Although the representations for the complete elliptic integrals given in (\ref{first_kind_period_a}) and (\ref{third_kind_period_a}) above are fairly standard ones (with relatively efficient implementations in \textsc{Mathematica}, for example), it is worthwhile to outline an alternative form for these integrals---which we hope has some promise to generalize beyond the elliptic case. In this appendix, we outline how these elliptic periods can be expressed in terms of Lauricella hypergeometric functions. (We refer the reader to e.g.~\cite{Brown:2019jng,Abreu:2019wzk,Abreu:2019xep} for some discussions on these functions in the context of Feynman integrals.) 

Consider first the elliptic period integral given in (\ref{first_kind_period_a}):
\eq{I^{(1)}_a\equivR\oint_{\Omega_a}\,\,\frac{\dbar\r\alpha}{y(\r\alpha)}=\frac{2\,i}{\pi\sqrt{r_{32}r_{41}}}K[\phi]\,,\label{I1a0_appendix}}
where $y^2(\r\alpha)\equivR(\r\alpha-r_1)(\r\alpha-r_2)(\r\alpha-r_3)(\r\alpha-r_4)$, $r_{ij}\equivR (r_i\mi r_j)$, and 
\eq{\phi\equivR\frac{(r_2{-}r_1)(r_3{-}r_4)}{(r_2{-}r_3)(r_1{-}r_4)}\equivL \frac{r_{21}r_{34}}{r_{23}r_{14}}.}
(We refer the reader to section \ref{subsec:elliptic_double_box_cut_of_sym} for our conventions regarding the ordering of the roots $r_i$.) This integral can be re-cast slightly by the change of variables 
\eq{\r{x}\equivR\frac{\r\alpha-r_1}{r_2-r_1}\,,}
upon which (\ref{I1a0_appendix}) becomes 
\eq{I^{(1)}_a\equivR\oint_{\Omega_a}\,\,\frac{\dbar\r\alpha}{y(\r\alpha)}=\frac{i}{\pi\sqrt{r_{32}r_{41}}}\int_0^1\!\!\!d\r{x}\frac{1}{\sqrt{\r{x}(1-\r{x})(1-\u{21}{31}\r{x})(1-\u{21}{41}\r{x})}},}
where we have introduced the shorthand
\eq{\u{ab}{cd}\equivR\frac{r_a-r_b}{r_c-r_d}\,.}
From the definition of the Lauricella hypergeometric function
\eq{\begin{split}
&F^{(n)}\left(a,b_{1},\dots,b_{n},c\Big|x_{1},\dots,x_{n}\right)\equivR\\
&\frac{\Gamma(c)}{\Gamma(a)\Gamma(c-a)}\int_{0}^{1}\!\!\!d\b{x}\,\,\, \b{x}^{a-1}(1-\b{x})^{c-a-1}(1-x_{1}\b{x})^{-b_{1}}\dots (1-x_{n}\b{x})^{-b_{n}}
\end{split}}
we have the identification
\eq{I^1_a=\frac{i\Gamma\left(\frac{1}{2}\right)^{2}}{\pi\sqrt{r_{31}r_{41}}}F^{(2)}\left(\frac{1}{2},\frac{1}{2},\frac{1}{2},1\Big|\u{21}{31},\u{21}{41}\right)\,.}

Applying the same transformation to the second fundamental period integral (\ref{third_kind_period_a}),
\eq{I^{(2)}_a\equivR\oint_{\Omega_a}\,\,\frac{\dbar\r\alpha\,\,y(p)}{(\r\alpha-p)y(\r\alpha)}\,,\label{I1b0_appendix}}
results in a representation 
\eq{I^{(2)}_a=\frac{-i\,y(p)\Gamma\left(\frac{1}{2}\right)^{2} }{\pi(p-r_{1})\sqrt{r_{31}r_{41}}}F^{(3)}\left(\frac{1}{2},\frac{1}{2},\frac{1}{2},1,1\Big|\u{21}{31},\u{21}{41},\u{21}{p1}\right).\label{I2a_appendix_lauricella}}

\newpage
\providecommand{\href}[2]{#2}\begingroup\raggedright\endgroup

\end{document}